\documentstyle[sprocl]{article}

\input epsf

\def\Journal#1#2#3#4{{#1} {\bf #2}, #3 (#4)}

\def\NCA{\em Nuovo Cimento}

\def\NPB{{\em Nucl. Phys.} B}
\def\PLB{{\em Phys. Lett.}  B}
\def\PRL{\em Phys. Rev. Lett.}
\def\PRD{{\em Phys. Rev.} D}
\def\ZPC{{\em Z. Phys.} C}

\def\st{\scriptstyle}
\def\sst{\scriptscriptstyle}
\def\mco{\multicolumn}
\def\epp{\epsilon^{\prime}}
\def\vep{\varepsilon}
\def\ra{\rightarrow}
\def\ppg{\pi^+\pi^-\gamma}
\def\vp{{\bf p}}
\def\ko{K^0}
\def\kb{\bar{K^0}}
\def\al{\alpha}
\def\ab{\bar{\alpha}}
\def\CPbar{\hbox{{\rm CP}\hskip-1.80em{/}}}%temp replacemt due to no font

\def\dalemb#1#2{{\vbox{\hrule height .#2pt
        \hbox{\vrule width.#2pt height#1pt \kern#1pt
                \vrule width.#2pt}
        \hrule height.#2pt}}}
\newenvironment{scalepic}[3]
  {\begin{center} \stepcounter{equation}
\def\@currentlabel{\theequation}      
   \label{#3}
   \begin{picture}(350,50)(-40,-25)%
   \put(0,10){\makebox(0,0)[rb]{#1}}
   \put(0,-10){\makebox(0,0)[rt]{#2}}
   \put(0,0){\makebox(-50,0){$\updownarrow$}}
   \put(0,0){\vector(1,0){300}}
   \put(356,0){\makebox(0,0)
} }
  { \end{picture} \end{center} }
\newcommand{\scaleitem}[3]
 { \put(#1,0){ \put(0,-5){\line(0,1){10}}
               \put(0,10){\makebox(0,0)[b]{#2}}
               \put(0,-10){\makebox(0,0)[t]{#3}} }}

\def\dalemb#1#2{{\vbox{\hrule height .#2pt
        \hbox{\vrule width.#2pt height#1pt \kern#1pt
                \vrule width.#2pt}
        \hrule height.#2pt}}}

 \def\bd{\begin{document}} \def\ed{\end{document}}
\def\ds{\documentstyle} \let\fr=\frac \let\bl=\bigl \let\br=\bigr
\let\Br=\Bigr \let\Bl=\Bigl 
\let\bm=\bibitem
\let\na=\nabla
\let\pa=\partial \let\ov=\overline
\def\ie{{\it i.e.\ }} 
\newcommand{\beq}{\begin{equation}}
\newcommand{\eeq}{\end{equation}}
\newcommand{\pr}{\paragraph{}}
\newcommand{\be}{\begin{equation}}
\newcommand{\ee}{\end{equation}}
\newcommand{\beba}{\begin{equation}\begin{array}{lcl}}
\newcommand{\eaee}{\end{array}\end{equation}}
\newcommand{\bea}{\begin{eqnarray}}
\newcommand{\eea}{\end{eqnarray}}
\newcommand{\ba}{\begin{array}}
\newcommand{\ea}{\end{array}}
\newcommand{\td}{\tilde}
\newcommand{\norsl}{\normalsize\sl}
\newcommand{\ns}{\normalsize}
\newcommand{\refs}[1]{(\ref{#1})}
\def\simlt{\mathrel{\lower2.5pt\vbox{\lineskip=0pt\baselineskip=0pt
           \hbox{$<$}\hbox{$\sim$}}}}
\def\simgt{\mathrel{\lower2.5pt\vbox{\lineskip=0pt\baselineskip=0pt
           \hbox{$>$}\hbox{$\sim$}}}}

\def\st{\scriptstyle}
\def\sst{\scriptscriptstyle}
\def\mco{\multicolumn}
\def\epp{\epsilon^{\prime}}
\def\vep{\varepsilon}
\def\ra{\rightarrow}
\def\ppg{\pi^+\pi^-\gamma}
\def\vp{{\bf p}}
\def\ko{K^0}
\def\kb{\bar{K^0}}
\def\al{\alpha}
\def\ab{\bar{\alpha}}
\def\be{\begin{equation}}
\def\ee{\end{equation}}
\def\bea{\begin{eqnarray}}
\def\eea{\end{eqnarray}}
\def\CPbar{\hbox{{\rm CP}\hskip-1.80em{/}}}
\begin{document}
\thispagestyle{empty}
\rightline{\normalsize\sf hep-ph/0007226}
\rightline{\normalsize CERN-TH/2000-125}
\rightline{\normalsize July 2000}
\vskip 1.0truecm
\centerline{\Large\bf Large Dimensions and String Physics in Future Colliders }
\vskip 1.truecm
\centerline{{\large\bf I. Antoniadis}\footnote{On leave from {\it Centre de
Physique  Th{\'e}orique, Ecole Polytechnique,
91128 Palaiseau, France}
(unit{\'e} mixte du CNRS et de l'EP, UMR 7644.)}
 and  
{\large\bf K. Benakli}}
\vskip .5truecm
%\centerline{\it On leave from Ecole Polytechnique, 91128 Palaiseau, France}
\vskip .5truecm
\centerline{{\it CERN Theory Division
 CH-1211, Gen{\`e}ve 23, Switzerland}}
\vskip .5truecm

\centerline{\bf\small ABSTRACT}
\vskip .4truecm

We review the status of low-scale string theories and large
extra-dimensions. After an overview on different string realizations, we
discuss some of the main important problems and we 
summarize present bounds on the size of possible 
extra-dimensions from collider experiments.

\hfill\break
\vfill\eject

\noindent
{\bf 1.}~~ Introduction\\
{\bf 2.}~~ Hiding extra dimensions\\
2.1~ Compactification on tori and Kaluza-Klein states\\
2.2~ Orbifolds and localized states\\
2.3~ Early motivation for large extra dimensions\\
{\bf 3.}~~ Low-scale strings\\
3.1~ Type I/I$^\prime$ string theory and D-branes\\
3.2~ Type II theories\\
3.3~ Heterotic string and  M-theory on $S^1/Z_2$``$\times$"Calabi-Yau\\
3.4~ Relation between Type I/I$^\prime$ and Type II with heterotic strings\\
{\bf 4.}~~ Theoretical implications\\
4.1~ U.V./I.R. correspondance\\
4.2~ Unification \\
4.3~ Supersymmetry breaking and scales hierarchy \\
4.4~ Electroweak symmetry breaking in TeV-scale strings\\
{\bf 5.}~ Scenarios for studies of experimental constraints\\
{\bf 6.}~ Extra-dimensions along the world brane: KK excitations of gauge
bosons\\
6.1~ Production at $e^+e^-$ colliders\\
6.2~ Production at hadron colliders\\
6.3~ High precision data low-energy bounds\\
6.4~ One extra dimension for other cases\\
6.5~ More than one extra dimension \\
{\bf 7.}~ Extra-dimensions transverse to the world brane: KK excitations
of gravitons\\
7.1~ Signals from missing energy experiments\\
7.2~ Gravity modification and sub-millimeter forces\\
{\bf 8.}~ Dimension-eight operators and  limits on the string scale\\
{\bf 9.}~ Conclusions\\
References
%$~$~~~~~~ References
\vfill
\section{Introduction}

In how many dimensions do we live? Could they be more than the four we
 are aware of? If so, why don't we see the other dimensions? Is there
 a way to detect them?. While the possibility of extra-dimensions has
 been considered by physicists  for long time, a compelling reason for
 their existence has arisen with  string theory. It  seems that a
 quantum theory of gravity requires that we live in more than  four
 dimensions, probably in ten or eleven dimensions.  The remaining
 (space-like) six or seven dimensions are hidden to us: observed
 particles do not propagate in them.  The theory  does not tell us yet
 why  four and only four have been  accessible to us. However, it
 predicts that this is only a low-energy effect: with increasing energy,
 particles which propagate in a higher dimensional space could be
 produced. What is the  value of the needed high energy scale? 
 could it be just close by, at reach of near  future experiments?

Another scale which appears in our attempts to answer the previous
questions is related to  the extended  nature of fundamental
objects.  It is the scale at which  internal degrees of freedom are
excited. In string theory this scale $M_s$  is related to the string
tension and sets the mass of the first heavy oscillation mode. The
point-like behavior of known particles as  observed at present
colliders allows to conclude that $M_s$ has to be higher than a
few hundred GeV. However to answer the question of what
energies should be reached before starting to probe this substructure of
the ``fundamental particles'', more precise determination of
experimental lower bounds on $M_s$ and understanding the assumptions
behind them is needed.

It is the aim of this review as to provide a short summary of the present 
status of research on  
extra-dimensions and string-like sub-structure of matter.

\section{ Hiding Extra-Dimensions }

\subsection{Compactification on Tori and Kaluza-Klein states}

There is a simple and elegant way to hide the extra-dimensions:
compactification. It is simple because it relies on an elementary
observation. Suppose that  the extra-dimensions form, at each point of
our four-dimensional space, a  $D$-dimensional torus of
volume $(2 \pi)^D  R_1 R_2 \cdots R_D$. 
The $(4+D)$-dimensional Poincare invariance is replaced by a four-dimensional
one times the symmetry group of the  $D$-dimensional
space which contains translations along the $D$ extra directions. The
$(4+D)$-dimensional momentum satisfies the mass-shell condition
$P_{(4+D)}^2 = p_0^2-p_1^2-p_2^2-p_3^2-\sum_i p_i^2 = m_0^2$ and looks from
the four-dimensional point of view as a (squared) mass $ M^2_{KK}=
p_0^2-p_1^2-p_2^2-p_3^2=m_0^2+\sum_i p_i^2$. Assuming periodicity of
the wave functions along each compact direction, one has $p_i = n_i/R_i$
which leads to:
\be
M^2_{KK}\equiv M^2_{\vec n} = m_0^2 +\frac {n_1^2}{R_1^2} +
\frac {n_2^2}{R_2^2}+ \cdots  +\frac {n_D^2}{R_D^2}\, ,
\label{KKdef}
\ee
with $m_0$ the higher-dimensional mass and $n_i$ non-negative
integers.  The states with $\sum_i n_i \neq 0$ are called Kaluza-Klein
(KK) states. It is clear that getting aware of the $i$th
extra-dimension  would require experiments that probe at least an
energy of the order of  $min(1/ R_i)$ with sizable couplings of the
KK  states to four-dimensional matter.

Let us discuss further some properties of the KK states that will be useful
for us below.  We parametrise the ``internal'' $D$-dimensional box by $y_i
\in [-\pi R_i, \pi R_i]$,  $i= 1,\cdots, D$ while  the four-dimensional
Minkowski spacetime is spanned by the coordinates $x^\mu$,   
$\mu= 0, \cdots, 3$. It is useful to choose for
the KK wave functions the basis:
\be
\Phi^{\alpha}_{{\vec n},{\vec e}} (x^\mu,y_i)= \Phi^{\alpha} (x^\mu) \,  
\prod_i \left[  
(1-e_i)  \cos (\frac {n_i y_i}{R_i}) + e_i \sin (\frac {n_i y_i}{R_i})
\right]\, ,
\label{KKmodes}
\ee 
where the vector ${\vec n}= (n_1, n_2, \cdots, n_D)$ gives  the
 energy of the state following Eq.~(\ref{KKdef}) while  $ {\vec e}
 =(e_1,\cdots, e_D)$ with $e_i = 0$  or 1 corresponds to a choice of 
cosine or sine dependence in the coordinate $y_i$, respectively. 
 The index ${\alpha}$ refers to other quantum numbers of $\Phi$.

\subsection{Orbifolds and localized states}
The simplest example of the models we will be using for getting
experimental bounds are  obtained by gauging the  $Z_2$ parity:
$y_i \rightarrow - y_i \,  {\rm mod} \,   2\pi R_i$. This leads
to compactification on segments of size $\pi R_i$. In general, the
consistency of  this ``orbifold'' projection  implies that
the $Z_2$ space parity  should be associated with a $Z_2$  action on 
the internal quantum numbers $\alpha$ of $\Phi$. As a result one has 
the following properties:

\begin{itemize}

\item Only states invariant under this $Z_2$ are kept while the others
are  projected out. There are two  classes of states left in the
theory: those for which $\Phi^{(even)} (x^\mu)$  is even under the $Z_2$
action and $e_i =0$ and those for which $\Phi^{(odd)} (x^\mu)$ is odd and
$e_i =1$. It is important to notice that the latter are not present as light
four-dimensional states   i.e. they have $\sum_i n_i \neq 0$ and thus
always correspond to higher KK states.

\item At the boundaries $y_i=0, \pi R$ fixed by the $Z_2$ action, 
new states $\Phi^{(loc)} (x^\mu)$, have to be
included. These ``twisted'' states are localized at the fixed points. They
can not propagate in the extra-dimension and thus have no KK excitations. 

\item The odd bulk states $\Phi^{(odd)} (x^\mu)$ ($e_i =1$) have a wave
function which vanishes  (the $\sin (\frac {n_i y_i}{R_i})$ in
Eq.~(\ref{KKdef} ) at the boundaries. Their coupling  to localized states
involves a derivative along $y_i$. For example three boson interactions of
the form $\partial_i \phi^{(odd))} \phi^{(loc)}\phi^{(loc)}$ can be
non-vanishing.

\item The even states, in contrast, can have non-derivative couplings to
localized states. The gauge couplings for instance are given by:
\be
g_n = {\sqrt{2}}\delta^{-{|\vec{\frac{n}{R}}|^2}/{M_s^2}} g
\label{coupling}
\ee
where $\delta >1$ is a model dependent number ($\delta=4$ in the case of
$Z_2$). The ${\sqrt{2}}$ comes from the relative normalization of
$\cos(\frac {n_i y_i}{R_i})$ wave function with respect to the zero mode
while the exponential damping is  a result of tree-level string computations
that we do not present here.

The exchange of KK states gives rise to an effective four-fermion operator:
\be
 {\bar \psi_1} {\psi_2} {\bar \psi_3} {\psi_4} 
\sum_{|\vec{n}|} \frac {g^2{(|\vec{n}|)}}{m_0^2 +
\frac{|\vec{n}|^2}{R^2}}\, .
\ee
The usual approximation of taking $g^2{(|\vec{n}|)}$ independent of
$|\vec{n}|$ fails for more than one dimension because the sum $\sum_{n_i}
\frac {1}{ n_1^2 +n_2^2 +...}$ becomes divergent. This divergence is
regularized by the exponential damping of Eq.~(\ref{coupling}). For
$D>1$ the result depends then on both parameters $R$ and
$M_s$. For $D=1$ the sum simplifies for large radius 
$M_s R \simgt 10$ as the sum converges rapidely and gives 
$ \frac {\pi^2} {3} g^2 R^2 {\bar \psi_1} {\psi_2} {\bar \psi_3} {\psi_4}$
which depends only on $R$.

\item Below we will be interested in string vacua where gauge degrees
of  freedom are localized on $(3+d_\parallel) +1$-dimensional subspaces:
$(3+d_\parallel)$-branes. From  the  point of view of 
 $(3+d_\parallel) +1$-dimensions  the gauge bosons behave as
``untwisted'' (not localized)  particles. In contrast, there are two
possible choices for light matter fields. In the first case,
 they arise from light modes 
of open strings with both ends on the $(3+d_\parallel)$-branes,
thus in their interactions they conserve momenta in the $d_\parallel$
directions. The second case are states that live on the intersection of the 
$(3+d_\parallel)$-branes with some other branes that do not contain the 
$d_\parallel$ directions in their worldvolume. These states are localized in 
the  $d_\parallel$-dimensional space and do not conserve the momenta in these directions. They 
have no KK excitations and behave as the $Z_2$ twisted (boundary)
states of heterotic strings on orbifolds.\footnote{In contrast to the
heterotic case open strings do not lead to $Z_N$ twisted matter with
$N>2$.} The boundary states couple to all KK-modes of gauge fields as described by Eq.~(\ref{coupling}). These couplings violate
obviously momentum conservation in the compact direction and make all
massive KK excitations unstable.
\end{itemize}

Use of compactification is an elegant way to hide extra-dimensions
because some of the quantum numbers and  interactions of the
elementary particles could be accounted to by the  topological and
geometrical properties of the internal space. For instance chirality,
number of families in the standard model, gauge and supersymmetry
breaking as well as as some  selection rules in the interactions of
light states  could be reproduced through judicious choice of more
complicated internal spaces.

\subsection{Early motivation for large extra-dimensions}

Attempts to construct a consistent theory for quantum gravity have
lead  only to one candidate: string theory. The only vacua of string
theory free of any pathologies are supersymmetric. Not being observed
in nature, supersymmetry should be broken.

In contrast to ordinary supergravity, where supersymmetry breaking can be
introduced at an arbitrary scale, through for instance the gravitino,
gaugini and other soft masses, in string theory this is not possible
(perturbatively). The only way to break supersymmetry at a scale
hierarchically smaller than the (heterotic) string scale is by
introducing a large compactification radius whose size is set by the
breaking scale. This has to be therefore of the order of a few TeV in
order to protect the gauge hierarchy. An explicit proof exists for
toroidal and fermionic constructions, although the result is believed to
apply to all compactifications~\cite{ablt,kp}. This is one of the very
few general predictions of perturbative (heterotic) string theory that
leads to the spectacular prediction of the possible existence of extra
dimensions accessible to future accelerators~\cite{ia}. The main
theoretical problem is though that the heterotic string coupling becomes necessarily strong.

The strong coupling problem can be understood from the effective field
theory point of view from the fact that at energies higher than the
compactification scale, the KK excitations of gauge bosons and other
Standard Model particles will start being produced and contribute to
various physical amplitudes. Their multiplicity turns very rapidly the
logarithmic evolution of gauge couplings into a power
dependence~\cite{tv}, invalidating the perturbative description, as
expected in a higher dimensional non-renormalizable gauge theory. A
possible way to avoid this problem is to impose conditions which prevent
the power corrections to low-energy couplings~\cite{ia}. For gauge
couplings, this implies the vanishing of the corresponding
$\beta$-functions, which is the case for instance when the KK modes are
organized in multiplets of $N=4$ supersymmetry, containing for every
massive spin-1 excitation, 2 Dirac fermions and 6 scalars. Examples of
such models are provided by orbifolds with no $N=2$ sectors with respect
to the large compact coordinate(s).

The simplest example of a one-dimensional orbifold is an interval of
length $\pi R$, or equivalently $S^1/Z_2$ with $Z_2$ the coordinate
inversion. The Hilbert space is composed of the untwisted sector,
obtained by the $Z_2$-projection of the toroidal states, and
of the twisted sector which is localized at the two end-points of the
interval, fixed under the $Z_2$ transformations. This sector is chiral
and can thus naturally contain quarks and leptons, while gauge fields
propagate in the (5d) bulk.

Similar conditions should be imposed to Yukawa's and in principle to
higher (non-renormalizable) effective couplings in order to ensure a soft
ultraviolet (UV) behavior above the compactification scale. We now know
that the problem of strong coupling can be addressed using string
S-dualities which invert the string coupling and relate a strongly
coupled theory with a weakly coupled one. For instance, as we
will discuss below, the strongly coupled heterotic theory with one large
dimension is described by a weakly coupled type IIB theory with a tension
at intermediate energies $(Rl_H)^{-1/2}\simeq 10^{11}$ GeV~\cite{ap}.
Furthermore, non-abelian gauge interactions emerge from tensionless
strings~\cite{w95} whose effective theory describes a higher-dimensional
non-trivial infrared fixed point of the renormalization group~\cite{sei}.
This theory incorporates all conditions to low-energy couplings that
guarantee a smooth UV behavior above the compactification scale. In
particular, one recovers that KK modes of gauge bosons form $N=4$
supermultiplets, while matter fields are localized in four dimensions. It
is remarkable that the main features of these models were captured
already in the context of the heterotic string despite its strong
coupling~\cite{ia}.

In the case of two or more large dimensions, the strongly coupled
heterotic string is described by a weakly coupled type IIA or type
I/I$^\prime$ theory~\cite{ap}. Moreover, the tension of the dual string
becomes of the order or even lower than the compactification scale. In
fact, as it will become clear in the following,  the string tension 
becomes an arbitrary parameter~\cite{w}. It
can be anywhere below the Planck scale and as low as a few TeV~\cite{l}.
The main advantage of having the string tension at the TeV, besides its
obvious experimental interest, is that it offers an automatic protection to
the gauge hierarchy, alternative to low-energy supersymmetry or
technicolor~\cite{add,aadd,ab}.

\section {Low-scale Strings}

In ten dimensions,  superstring theory has two parameters: a mass (or
length) scale $M_s$ ($l_s=M_s^{-1}$), and a dimensionless string
coupling $g_s$ given by the vacuum expectation value (VEV) of the
dilaton field $e^{<\phi>}=g_s$ on which we impose the
weakly coupled  condition $g_s<1$. Compactification to lower
dimensions introduces other  parameters describing for instance
volumes and shapes of the internal space.  The $D$-dimensional
compactification volume $V_D$ will always be chosen to be bigger than
unity in string units, $V_D \ge l_s^D$. This choice can always be done
by appropriate  T-duality transformations which inverts
the compactification radius. To illustrate this duality let us consider a string vacuum with a $d_\parallel$-brane on which the standard model gauge bosons are localized. There are three type of strings:

\begin {itemize}

\item Closed strings have masses given by
\be
M^2_{closed}= \sum_{i=1}^{d_\parallel} \frac {n_i^2}{R_{\parallel i}^2} 
+\sum_{j=1}^{d_\perp = 9-d_\parallel}\frac {n_j^2}{R_{\perp i}^2}
+\sum_{i=1}^{d_\parallel} \frac {m_i^2R_{\parallel i}^2}{l_s^4} +
\sum_{j=1}^{d_\perp = 9-d_\parallel}\frac {m_j^2R_{\perp j}^2}{l_s^4}+ 
\frac {N} {l_s^2}\, ,
\label{mclosed1}
\ee

\item open strings with both ends on the $d_\parallel$-brane with masses

\be
M^2_{DD}= \sum_{i=1}^{d_\parallel} \frac {n_i^2}{R_{\parallel i}^2} +
\sum_{j=1}^{d_\perp = 9-d_\parallel}\frac {m_j^2R_{\perp j}^2}{l_s^4}+ 
\frac {N} {l_s^2}\, ,
\label{mopen2}
\ee

\item open strings with one ends on the $d_\parallel$-brane and another 
on a $d'_\parallel$-brane intersecting along 
$d_\parallel \bigcap d'_\parallel$ dimensions, for which the mass formula 
reads
\be
M^2_{DD'}= \sum_{i \in d_\parallel \bigcap d'_\parallel} \frac {n_i^2}{R_{\parallel i}^2} +
\sum_{j \in 9-(d_\parallel \bigcup d'_\parallel) }\frac {m_i^2R_{\perp i}^2}{l_s^4}+ 
\frac {N} {l_s^2}\, ,
\label{mopen3}
\ee
where $n_i, m_i$ and $N$ are integer numbers. Note that the later have 
neither KK excitations 
($p={m\over R}$) nor  winding modes ($w={nR\over l_s^2}$) along the directions 
 $(d_\parallel \bigcup d'_\parallel) - 
(d_\parallel \bigcap d'_\parallel)$ in which they are localized. 

\end{itemize}

 T-duality  not only exchanges
Kaluza-Klein (KK) momenta $p={m\over R}$ with string winding modes
$w={nR\over l_s^2}$, but also rescales the string
coupling: 

\be R\to{l_s^2\over R}\qquad g_s \to g_s{l_s\over R}\, ,
\label{Tdual}
\ee
so that the lower-dimensional coupling $g_s\sqrt{l_s/R}$ remains
invariant. When $R$ is smaller than the string scale, the winding modes
become very light, while T-duality trades them as KK momenta in terms of
the dual radius ${\tilde R}\equiv l_s^2/R$. The enhancement of the string
coupling is then due to their multiplicity which diverges in the limit
$R\to 0$ (or ${\tilde R}\to\infty$).

Upon compactification in $D=4$ dimensions, these parameters determine
the values at the string scale of the four-dimensional (4d) Planck
mass (or length) $M_p$ ($l_p=M_p^{-1}$) and  gauge coupling $g_{YM}$
that for phenomenological purposes should have  the correct strength
magnitude.  For instance, generically the four-dimensional Planck mass can be
 expressed as:
\be
M^2_{pl}  \equiv  f_{pl}  \frac {(M_s^6 V_6)} {g_s^2} M_s^2\, ,
\label{newtonH}
\ee
where $V_6$ is the six-dimensional internal volume felt by
gravitational interactions while the four-dimensional gauge coupling can be 
written as
\be
\frac {1}{g_{YM}^2} \equiv f_{YM} \frac {(M_s^d V_d)}{g_s^q}\, ,
\label{coupH}
\ee
where $V_d$ is the $d$-dimensional internal volume felt by gauge
interactions, and the coefficients $f_{pl}$, $f_{YM}$ have been computed
for known  classical string vacua. In the lowest order approximation,
they are moduli-independent ${\cal O}(1)$ constants \footnote{ Below we will often simplify the discussion by  taking $f_{pl} =f_{YM} =1$}.

In the past, weakly coupled heterotic strings were providing the
most promising framework for phenomenological applications. In this
case, the standard model was considered as descending from the
ten-dimensional $E_8$ gauge symmetry, and we have $V_d =V_6$, $d=6$ and
$q=2$. Taking the ratio of the two equations, one finds $ \frac
{M_s^2} {M^2_{pl}}= \frac { f_{YM}}{f_{pl}} g_{YM}^2   \sim g_{YM}^2$.
Requiring $g_{YM} \sim {\cal O}(1)$, it was concluded that both the 
string scale $M_s$
and the compactification scale $R^{-1} \equiv V_6^{-1/6}$ had to lie
just below the Planck scale, at energies $\sim 10^{18}$GeV far out of
reach of any near future experiment~\cite{ia,add}.

The situation changed during recent years when it was discovered
that string theory provides classical solutions (vacua) where gauge degrees
of freedom live on subspaces i.e. $d < D$ along with the possibility of 
$p \neq q$. For instance, while $D=6$ and $p=2$, $(d,q)=(d,1)$ in type I
and $(d,q)=(2,0)$ in type II or weakly coupled heterotic strings with small
instantons. In these cases, it is an easy exercise to check that both the
string and compactification scales can be made arbitrarily low.

The possibility of decreasing the string scale offers new insights on
the physics beyond the standard model. For instance, a string scale at
energies as low as TeV, would in addition to the plethora of
experimental signatures,  provides a  solution to the
problem of gauge hierarchy alternative to supersymmetry or technicolor. The
hierarchy in gauge symmetry versus fundamental (cut-off) scales is
then nullified as the two are of the same order
~\cite{add,aadd}. Another possibility \cite{int,burg} 
is an intermediate scale which
then identifies the string scale with natural scales where some new
physics is expected, as for instance the scale of supersymmetry
breaking in a hidden sector, the Peccei-Quinn axion physics, the 
neutrino see-saw scale etc. For instance, in a generic brane configuration, 
there might be a non-supersymmetric brane (as an anti-brane) which is located 
far away from the supersymmetric brane on which the standard model fields 
are localized. In this case supersymmetry is broken on the far brane at $M_s$ and if communicated through gravity, the scale of supersymmetry breaking on our
brane will be of order $M_s^2/M_{pl}$. Requiring the latter to be in the TeV
range implies a string scale at intermediate energies.

 We review below the different possible 
realizations of low scale string theories.

\subsection {Type I/I$^\prime$ string theory and D-branes}

 Type
I/I$^\prime$ is a ten-dimensional 
theory of closed and open unoriented strings. Closed
strings describe gravity, while gauge interactions are described by open
strings whose ends are confined to propagate on $p$-dimensional 
sub-spaces defined as 
D$p$-branes. The internal space has 6  compactified dimensions, $p-3$
longitudinal and $9-p$ transverse to the D$p$-brane.

The gauge and  gravitational interactions appear at
different order in string loops perturbation theory, leading to different powers of $g_s$ in the corresponding effective action:
\be
S_{I}=\int d^{10}x \frac{1}{g_s^2 l_s^8} {\cal R} + 
\int d^{p+1}x \frac{1}{g_s l_s^{p-3}} F^2\, ,
\label{StypeI}
\ee
The $1/g_s$ factor in front of the gauge kinetic terms corresponds to
the lowest order open string diagram represented  by a disk. 

Upon compactification in four dimensions, the Planck length and gauge
couplings are given to leading order by
\begin{equation}
\frac{1}{l_P^2}=\frac{V_\parallel V_\perp}{g_s^2 l_s^8}\ ,\qquad
\frac{1}{g_{YM}^2}=\frac{V_\parallel}{g_s l_s^{p-3}}\, ,
\label{I}
\end{equation}
where $V_\parallel$ ($V_\perp$) denotes the compactification volume 
longitudinal (transverse) to the $Dp$-brane. From the second relation
above, it follows that the requirements of weak coupling $g_{YM} \sim {\cal O}(1)$, $g_s<1$
imply that the size of the longitudinal space must be of order of the
string length ($V_\parallel\sim l_s^{p-3}$), while the transverse volume
$V_\perp$ remains unrestricted. Using the longitudinal volume in string units 
$v_\parallel\simgt 1$, and assuming an isotropic transverse space of $n=9-p$ compact 
dimensions of radius $R_\perp$, we can rewrite these realtions as:
\begin{equation}
M_P^2=\frac{1}{g_{YM}^4 v_\parallel}M_s^{2+n}R_\perp^n\ ,\qquad
g_s =g_{YM}^2 v_\parallel\, .
\label{treei}
\end{equation}

From the relations (\ref{treei}), it follows that the type I/I$^\prime$
string scale can be chosen hierarchically smaller than the Planck mass at
the expense of introducing extra large transverse dimensions that
are felt only by the gravitationally interacting light states, 
while keeping the string coupling
weak~\cite{aadd}. The weakness of 4d gravity compared to gauge interactions
 (ratio $M_W/M_P$) is then
attributed to the largeness of the transverse space $R_\perp/l_s$.

An important property of these models is that gravity becomes
$(4+n)$-dimensional  with a strength comparable to those of gauge
interactions at the string scale. The first relation of
eq.(\ref{treei}) can be understood as a consequence of the
$(4+n)$-dimensional Gauss law for gravity, with 
\be
G_N^{(4+n)}=g_{YM}^4 l_s^{2+n}v_\parallel
\label{GN}
\ee
the Newton's constant in $4+n$ dimensions.

Taking the type I string scale $M_s$ to be at 1 TeV,
one finds a size for the transverse dimensions $R_\perp$ varying from
$10^8$ km, .1 mm (10$^{-3}$ eV), down  to .1 fermi (10 MeV) for $n=1,2$,
or 6 large dimensions, respectively. This shows that while $d_\perp =1$ is 
excluded, $d_\perp \geq 2$ are allowed by present experimental bounds
on gravitational forces\cite{price}.

\subsection  {Type II theories}

We proceed now with discussion of the relations (\ref{newtonH}) and 
(\ref{coupH}) for the case of models derived from compactifications of 
Type II strings. For simplicity, we shall restrict ourselves to 
four-dimensional compactifications
of type II on $K3\times T^2$, yielding $N=4$ supersymmetry. 
 Calabi-Yau manifolds that lead to 
$N=2$ supersymmetry can be obtained by replacing $T^2$ by a ``base"
two-sphere over which $K3$ varies while more interesting
phenomenological models with $N=1$ supersymmetry can be obtained by a
freely acting orbifold, although the most general $N=1$
compactifications would require F-theory on Calabi-Yau fourfolds.

In type IIA non-abelian gauge symmetries arise in six dimensions from
D2-branes wrapped around non-trivial vanishing 2-cycles of a singular
$K3$. The gauge kinetic terms are independent of the string coupling
$g_s$ and the corresponding effective action is:
\be
S_{IIA}=\int d^{10}x \frac{1}{g_s^2 l_s^8} {\cal R} + 
\int d^6 x {1\over l_s^2} F^2\, .
\label{SIIA}
\ee 
Upon compactification on a two-torus $T^2$ of size $V_{T^2}$ to
four dimensions, the gauge couplings are determined by $V_{T^2}$,
while the Planck mass is also controlled by $V_{K3}$ and $g_s$: 
\be
\frac{1}{g_{YM}^2}={V_{T^2}\over l_s^2} \qquad\qquad
\frac{1}{l_P^2}=\frac{V_{T_2}V_{K3} }{g_s^2 l_s^8} ={V_{K3}\over g_s^2
l_s^6 }{1\over g_{YM}^2}\, .
\label{IIA}
\ee

Therefore the area of $T^2$ should be of order $l_s^2$, while both
$g_s$ and $V_{K3}$ can be used to separate the Planck mass from the
string scale~\cite{l,ap}: 
\be 
M_s= M_P g_{YM} {g_s l_s^2\over\sqrt{V_{K3}}}\, ,
\label{IIA2}
\ee
Taking $M_s \sim M_W$, with $M_W$ the weak scale,
 the hierarchy between the electroweak and the Planck
scales  could be now obtained with a choice of string-size internal
manifold and an ultra-weak coupling $g_s=10^{-14}$~\cite{ap}. As a result, 
gravity remains weak even at the string scale where  the corresponding string
interactions are suppressed by the tiny string coupling, or equivalently
by the 4d Planck mass. The main observable effects in particle
accelerators are the production of KK excitations along the two
TeV dimensions of $T^2$ with gauge interactions.

In a way similar to the case of Type I/I' strings,  one can instead produce  
a  hierarchy of scales$M_s/M_P$ by keeping $g_s$ of order unity and 
allowing  some of 
the $K3$ (transverse) directions to be
large. This corresponds to 
$V_{K3}/l_s^4\sim 10^{28}$, implying a fermi size for the four $K3$
compact dimensions. Alternatively, one could play with both parameters 
$g_s$ and $V_{K3}$.

An intersting possibility to mention is that 
it is possible to  satisfy  eq.(\ref{IIA}) while taking one direction 
much bigger 
than the string scale and the other much smaller. For instance, in the case of a rectangular
torus of radii $r$ and $R$, $V_{T^2}=rR\sim l_s^2$ with
$r\gg l_s\gg R$. This can be treated by performing a T-duality
(\ref{Tdual}) along $R$ to type IIB: $R\to l_s^2/R$ and
$g_s\to{\tilde g_s}=g_s l_s/R$ with
$l_s={\tilde l_s}$. One thus obtains:
\be
\frac{1}{g_{YM}^2}={r\over R} \qquad\qquad
\frac{1}{l_P^2}=\frac{V_{T_2} V_{K3}}{{\tilde g_s}^2 {\tilde l_s}^8}
={R^2 V_{K3}\over  {\tilde g_s}^2 {\tilde l_s}^6}{1\over g_{YM}^2}\, .
\label{IIB}
\ee
which shows that the gauge couplings are now determined by the ratio of
the two radii, or in general by the shape of $T^2$, while the Planck mass
is controlled by its size.

Since $T^2$ is felt by gauge interactions, its size cannot be larger
than ${\cal O}({\rm TeV}^{-1})$ implying that in a scenario where
$R\gg {\tilde l_s}$, the type IIB string scale should be much larger
than TeV. The condition of weakly coupled ten (and six) dimensional
type II theory implies ${\tilde M_s}\simlt\sqrt{M_{Pl}/R}$, so that the
largest value for the string tension, when $R\sim 1{\rm TeV}^{-1}$, is
an intermediate scale $\sim 10^{11}$ GeV when the string coupling is
of order unity.  In the energy  range between the KK scale $1/R$ and
the type IIB string scale, one has an effective 6d theory without
gravity at a non-trivial superconformal fixed point described by a
tensionless string~\cite{w95,sei}. This is because in type IIB gauge
symmetries still arise non-perturbatively from vanishing 2-cycles of
$K3$, but take the form of tensionless strings in 6 dimensions, given by
D3-branes wrapped on the vanishing cycles. Only after further
compactification does this theory reduce to a standard gauge theory,
whose coupling involves the shape rather than the volume of the
two-torus, as described above. Since the type IIB coupling is of order
unity, gravity becomes strong at the type IIB string scale and the main
experimental signals at TeV energies are similar to those of type IIA
models with tiny string coupling i.e. production of KK excitations of gauge
degrees of freedom.

\subsection {Heterotic string and  M-theory on $S^1/Z_2$``$\times$"Calabi-Yau}

As we have stated in the begining of this section, the weakly coupled perturbative heterotic string vacua predict that $ \frac
{M_s^2} {M^2_{pl}}= \frac { f_{YM}}{f_{pl}} g_{YM}^2   \sim g_{YM}^2$
leading to both the string and compactification scales in the energy ranges 
$\sim 10^{18}$GeV far out of
reach of any near future experiment. 

\subsubsection {M-theory on $S^1/Z_2$``$\times$"Calabi-Yau}

Let us first discuss the possibility of going to the strong coupling
limit.  The strong coupling limit of $SO(32)$ heterotic strings is
described by  type I strings and we have seen that they allow for
arbitrarely low scales.  The strong coupling dual of $E_8 \times E_8$
heterotic strings is described  by the eleven dimensional M-theory on
an orbifold $S^1/Z_2$ of size $\pi \rho$ \cite{hw,witten}. Gauge
fields and  matter live on the two ten-dimensional boundaries while
gravitons propagate in the eleven-dimensional bulk.

A four-dimensional theory can be obtained by a further compactification 
on a Calabi-Yau 
manifold.  Following \cite{witten} one may solve the equations of motion for such
 configuration as a perturbative expansion  in the dimensionless
 parameter ${\rho  M_{11}^{-3}/V^{2/3}}$. At higher orders in this
 expansion,  the  factorization in a product $S^1/Z_2 \times CY$   is
 lost. The volume of the Calabi-Yau space  becomes  a function of  the
 coordinate parametrizing the $S^1/Z_2$ segment. More precisely, the
 volumes of $CY$ seen by the observable sector\footnote{ We will use
 the subscripts $o$ for parameters of the observable sector and $h$
 for those of the hidden sector.} $V_{o}$ and the one on the hidden
 wall $V_{h}$  are given by:   \be   V_{o}\equiv V(0) = V \left(1 + \left(\frac
 {\pi}{2}\right)^{4/3} a_o {\rho  M_{11}^{-3}\over V^{2/3}}\right)
\label{volo}
\ee  and  
\be  V_{h} \equiv V(\pi \rho) = V \left(1 + \left( \frac {\pi}{2}\right)^{4/3}
a_h {\rho M_{11}^{-3}\over V^{2/3}} \right)
\label{volh}
\ee
where now $V$ is the (constant) lowest order value for the volume of
the  Calabi-Yau manifold and $a_{o,h}$ are model-dependent constants
\cite{nse}. Roughly speaking $a_{o,h}$ count the proportion of
instantons and five-branes on each wall. The coefficients $a_o$ and $a_h$ are given by:

\be a_{o,h} =  \int_{CY} \omega \wedge \frac{ tr(F_{o,h} \wedge F_{o,h}) -
\frac{1}{2} tr(R \wedge R)}{8 \pi^2}
\ee
where $\omega$ is the Kahler two-form of the  Calabi-Yau. The Newton constant is given by: 

\be G_N  = {1\over 16 \pi^2}{1\over M_{11}^9  \rho \langle V\rangle} \,,
\ee
with $\langle V\rangle$ the average volume of the Calabi-Yau space
on the eleven dimensional segment.   The gauge couplings are given by:

\beq 
\alpha_{o,h} = (4\pi)^{2/3} \frac { 1}{ f_{o,h}  M_{11}^6 
 V_{o,h} }\, .
\eeq
where the constant $f_o$ ($f_h$) is a
ratio of  normalization of the traces of adjoint representation of
$G_o$ ($G_h$)  compare to $E_8$ case. In the absence of 5-branes, one obtains that 
$a_o = -a_h$ and $\langle V\rangle =V$ . Explicit  computations\cite{nse}
 show that $a_o$ can be either positive, zero or negative.

\vskip 0.5cm $\bullet$ {\em Case $a_o > 0 \rightarrow$  $M_{11} \simgt
10^{16}$ GeV}

Compactifications with standard embedding of the gauge connection fall
in this category (see \cite{witten}).  In these models there is an
upper limit on the size of  the $S^1/Z_2$ segment above which the
hidden sector gauge coupling blows up. If the observable sector
coupling constant is of order unity the corresponding lower
bound on the M-theory scale $M_{11}$ is of order  $10^{16}$ GeV.

$\bullet$ {\em Case $a_o = a_h =0 \rightarrow $  $M_{11}
\simgt 10^{7}$ GeV  }

 This can be obtained for example in symmetric embedding.
In this case the only  upper limit on $\rho$ is from experiments on
modification of the Newtonian force at distances of  $\rho \simgt $ mm.
 Using  $\langle V\rangle = V_o$ and $\alpha_{o}
\sim 1/10$ one obtained a lower  bound on  limit $M_{11}$ of the order
of $4\times 10^{7}$ GeV.

 $\bullet$ {\em Case $a_o <0 \rightarrow$  $M_{11} \simgt$
TeV with $\rho^{-1} \ll $ TeV}

The possibility of  $a_o <0$  arises in the
non-standard embedding in \cite{nse}.   In this
scenario, as $\rho$ increases the volume of the internal space on the
observable wall  is fixed as to fit the desired value of $\alpha_{o}$
while  the volume on  the other end of the segment increases leading
to smaller values of the corresponding coupling constant.  Typically,
$\langle V\rangle \sim \frac {V_h}{2} \gg V_o$ for  large  values of
the radius $\rho$.  Given a value of $M_{11}$ both $V_o$ and $\rho
\langle V\rangle$ can be tuned to fit the value of $\alpha_{o}$ and
$M_{Pl}$.

For an $M$-theory scale at TeV one finds that seven dimensions have to be anisotropically large: $\rho^{-1} \sim 20$ eV while $\langle V\rangle^{-1/6} \sim 2$ GeV.
In this case of non-standard embedding the  hidden observer living on 
the other wall could see
the new longitudinal 
dimensions at energies ( e.g. GeV) much before the observers
on our wall (TeV).  At energies of 
the order of $GeV$ the
states in the bulk are not anymore the plane waves  Kaluza-Klein
states. Instead, one expects heavier modes localized on our side of
the universe which decay to lighter massive modes  localized near the
other  wall before the latter decay to hidden matter.

\subsubsection {Small instantons}

Consider first the case of the 
weakly  coupled  $SO(32)$ heterotic string theory compactified
on a $K3$ leading to $N=1$ supersymmetry in six dimensions. Witten argued \cite{SO32} that at the singularity, 
associated with a collapse
of $k$ instantons at the same point in $K3$, a new $Sp(k)$ gauge symmetry appears.
In addition   
massless hypermultiplets appear. They consist of $({\bf 32},{\bf 2k})$
of the $SO(32) \times Sp(k)$ gauge group and a massless hypermultiplet in the antisymmetric representation of $Sp(k)$, which is a singlet of $SO(32)$.
The six-dimensional gauge coupling of the small instanton
sector at the heterotic side is given by
$g_{SI}^2=(2\pi)^3 l_H^2$. Further compactification to 
four dimensions, by using a  fibration of $K3$
over a $P^1$ base, leads to a  $N=1$ supersymmetric theory with a 
gauge coupling 
\beq
\alpha_{Sp} \ =\ \frac{2 \pi^2 l_H^2}{V_{P^1}} \ ,
\label{sp}
\eeq
where $V_{P^1}$ is the volume of the base.

The configuration where one identifies the standard model gauge group with 
the small instantons gauge sector will allow us to consider arbitrary low 
 heterotic $SO(32)$ string scale\cite{yaron}.

There are  dimensionless expansion parameters in the system that we require
to be small.
The first is the expansion parameter of the perturbative string description
in ten dimensions ${\lambda_H^2}/{\left( 2 \pi \right)^5}$ \cite{Kap}: 
\beq
\frac {\lambda_H^2}{\left( 2 \pi \right)^5} =\frac {2}{\pi^4 } 
\frac { l_p^2}{ l_H^2}\frac {\left< V_{K3} V_{P^1}\right>}{\lambda_H^6} \ ,
\label{st}
\eeq
which we require to be smaller
than 1 in order for the heterotic string to be weakly coupled
in space-time.
The second parameter is $\alpha_{Sp}$ in (\ref{sp}), which
we require to be smaller
than 1 in order for the new gauge symmetry to be weakly coupled. These are satisfied by choosing: (i) $K3$ such that  $\frac {l_H^4}{\left< V_{K3}\right>} < 1$ 
so that the small
instanton picture is valid (ii) 
$\left< V_{K3} V_{P^1}\right> \sim \left< V_{K3}\right> \left<  V_{P^1}\right> $ and $\alpha_{Sp} \sim \frac{l_H^2}{V_{P^1}}$ 
 small guarantee that $\frac {l_H^6}{\left< V_{K3} V_{P^1}\right>}$
is small too (iii) $\frac { l_p^2}{ l_H^2}$ be small
 in order for $\frac {\lambda_H^2}{\left( 2 \pi \right)^5}$
to be small, namely a 
weakness of
gravitational interactions is consistent with the weakly coupled description.

We can view the weakness of
gravitational interactions as arising either from a large $K3$ volume
or from a very small string coupling constant.  For instance, taking $\alpha_{Sp} \sim 1/10$ 
as a rough estimate, the first possibility
 arises, with a choice:\\ 
$ \lambda_H \sim 1$ \, and \, $\left< V_{K3}\right>^{1/4}\sim 10,\ \,  10^{3},\ \,  
10^{6} \,  l_H$ \, 
for \, $l_H^{-1} \sim 10^{16},\ \,  10^{11},\ \, 10^{4}$ GeV respectively.
The second possibility arises, with a choice:\\
$\left< V_{K3}\right>^{1/4} \sim  {\rm few}~~l_H$ \, and \, $ \lambda_H \sim 10^{-1},\,  10^{-6},\, 10^{-13}$,  for  $l_H^{-1} \sim \, 10^{16}, \ \,  10^{11}, \ \,  10^{4}$ GeV 
respectively.

If $\lambda_H$ is chosen to be very small, at energies below the string scale,
 the unbroken part of the
 $SO(32)$ symmetry is very weakly coupled and it is seen from the $Sp(k)$ side as a non-abelian ``global'' symmetry. Such kinds of symmetries can
 be useful for phenomenological issues such as forbidding
operators that could lead to proton decay or other exotic processes. On the other hand the gravitational interactions are still weak at the string scale. The main experimental signature would be the observation of effects due to the Kaluza--Klein modes of $P^1$. If one instead explains the weakness
of gravitational interactions by a large $K3$ volume (as in type I scenarios)
then at energies of order $l_H^{-1}$ the $SO(32)$ symmetry coupling is of the 
same order as the one of $Sp(k)$ and cannot be viewed as a global symmetry. This is due to the sum of the contributions from the Kaluza--Klein states propagating in the $K3$.
Moreover at the string scale the gravitational interactions are now of the same strength as  the gauge ones.

One could instead shrink instantons at ADE singularities
of $K3$ \cite{aspin,BI,Intri}.
The gauge groups are then products of the classical gauge groups
$\prod_{i,j,k} SO(n_i) \times Sp(m_j) \times U(l_k)$ arranged according to 
quiver (moose)  diagrams related to the extended Dynkin diagrams of the ADE groups.

Consider now the case of $E_8\times E_8$ heterotic string compactified 
on a $K3$ fibration
over a $P^1$ base, 
in the adiabatic limit.
Denote by $n_1,n_2$ the instanton numbers of the two
$E_8$ groups.
We have to choose the gauge bundle with 
$n_1+n_2=24$.
When we shrink some of the instantons to zero size we do not get a new gauge symmetry
in six dimensions. Instead, we get massless tensor multiplets and hypermultiplets
in six dimensions \cite{SW,GH}. 
The six-dimensional tensor multiplet contains a 2-form field $B_{\mu\nu}$
which is self-dual $dB =*dB$. 
In the dual picture of M-theory compactified on
$S^1/Z_2$, this process is viewed as  placing M5-branes near one of the $E_8$ walls.
There are tensionless strings that arise from membranes stretched between the
M5-branes and the $E_8$ wall and couple to $B$.
When we reduce on $P^1$ the tensor multiplets do  not give rise to gauge fields
but rather to matter multiplets. This is due to the fact that there are no 1-forms
$\omega$ on $P^1$, which otherwise would enable us to decompose
$dB = F \wedge \omega$ and obtain the gauge field strength $F$.

We can however obtain vectors fields in six dimensions
and a large class of gauge groups and matter content by shrinking $E_8$ instantons 
at ADE singularities \cite{aspin}.
For instance, if we shrink $k$ instantons at $A_{n-1}$ singularity
we get a gauge group $\prod_{i=2}^{n-1}SU(i)\times SU(n)^{k-2n+1}\times
 \prod_{j=2}^{n-1}SU(j)$
with bi-fundamental matter.
The six-dimensional gauge couplings of these gauge groups is determined by vacuum
expectation values (vev's) $\left < \phi \right >$ 
of scalars in particular tensor multiplets \cite{aspin}.
These
 scalars in six dimensions have dimension two 
and we can choose  vev's $\left < \phi \right > \sim 1/l_H^2$.
Upon reduction on $P^1$ we can identically repeat 
the discussion in the previous section for the weakly coupled heterotic 
strings case \cite{yaron}. For the Ho\v rava--Witten  compactifications an arbitrarily low scale can be obtained by taking  all or some of the five 
dimensions transverse to the M5-brane large.

\subsection{Relation type I/I$^\prime$ and type II -- heterotic}

The type I/I$^\prime$ and type II models discussed above 
describe particular strongly coupled heterotic vacua with large
dimensions~\cite{aq,ap}. Let us first consider the heterotic
string compactified on a 6d manifold with $k$ large dimensions of radius 
$R\gg l_H$ and $6-k$ string-size dimensions. One can show that for $k\ge 4$ it
has a perturbative type I$^\prime$ description~\cite{ap}. 

In ten dimensions, heterotic and type I theories are related by an
S-duality:
\be
\lambda_I={1\over\lambda_H}\qquad\qquad l_I=\lambda_H^{1/2}l_H\, ,
\label{het-I}
\ee
which can be obtained for instance by comparing the heterotic case:
\be
M_H=g_{YM}M_{pl}\qquad\qquad \lambda_H=g_{YM}{\sqrt{V}\over l_H^3}\, .
\label{het}
\ee
with the case of 9-branes ($p=9$, $V_\perp=1$,
$V_\parallel=V$ in eq.~(\ref{I})). Using from eq.(\ref{het}) that 
$\lambda_H\sim (R/l_H)^{k/2}$, one finds
\be
\lambda_I\sim\left({R\over l_H}\right)^{-k/2}\qquad\qquad
l_I\sim\left({R\over l_H}\right)^{k/4}l_H\, .
\ee
It follows that the type I scale $M_I$ appears as a non-perturbative
threshold in the heterotic string at energies much lower than
$M_H$~\cite{Kap}. For $k<4$, it appears at intermediate energies
$R^{-1}<M_I<M_H$, for $k=4$, it becomes of the order of the
compactification scale $M_I\sim R^{-1}$, while for $k>4$, it appears at
lower energies $M_I<R^{-1}$~\cite{aq}. Moreover, since $\lambda_I\ll 1$,
one would naively think that weakly coupled type I theory could describe
the heterotic string with any number $k\ge 1$ of large dimensions.
However, this is not true because there are always some dimensions
smaller than the type I size ($6-k$ for $k<4$ and 6 for $k>4$) and one
has to perform T-dualities (\ref{Tdual}) in order to account for the
multiplicity of light winding modes in the closed string sector, as we
discussed in eq.~(\ref{Tdual}). Note that open strings have no winding
modes along longitudinal dimensions and no KK momenta along transverse
directions. The T-dualities have two effects: (i) they transform the
corresponding longitudinal directions to transverse ones by exchanging KK
momenta with winding modes, and (ii) they increase the string coupling
according to eq.(\ref{Tdual}) and therefore it is not clear that type
I$^\prime$ theory remains weakly coupled.

Indeed for $k<4$, after performing $6-k$ T-dualities on the heterotic
size dimensions, with respect to the type I scale, one obtains a type
I$^\prime$ theory with D($3+k$)-branes but strong coupling:
\be
l_H\to{\tilde l}_H\!=\!{l_I^2\over l_H}\!\sim\!
\left({R\over l_H}\right)^{k/2}l_H
\qquad \lambda_I\to{\tilde\lambda}_I\!=\!
\lambda_I\left({l_I\over l_H}\right)^{6-k}\!\sim\!
\left({R\over l_H}\right)^{k(4-k)/4}\!\!\gg\!\! 1\, .
\label{kl4}
\ee
For $k\ge 4$, we must perform T-dualities in all six internal
directions.\footnote{The case $k=4$ can be treated in the same way, since
there are 4 dimensions that have type I string size and remain inert
under T-duality.} As a result, the type I$^\prime$ theory has D3-branes
with $6-k$ transverse dimensions of radius ${\tilde l}_H$ given in
eq.(\ref{kl4}) and $k$ transverse dimensions of radius 
${\tilde R}=l_I^2/R\sim (R/l_H)^{k/2-1}$, while its coupling remains weak
(of order unity):
\be
\lambda_I\to{\tilde\lambda}_I=\lambda_I
\left({l_I\over l_H}\right)^{6-k}\left({l_I\over R}\right)^k\sim 1\, .
\ee

It follows that the type I$^\prime$ theory with $n$ extra-large
transverse dimensions offers a weakly coupled dual description for the
heterotic string with $k=4,5,6$ large dimensions~\cite{ap}. $k=4$ is
described by $n=2$, $k=6$ (for $SO(32)$ gauge group) is described by
$n=6$, while for $n=5$ one finds a type I$^\prime$ model with 5 large
transverse dimensions and one extra-large. The case $k=4$ is particularly
interesting: the heterotic string with 4 large dimensions, say at a TeV,
is described by a perturbative type I$^\prime$ theory with the string
scale at the TeV and 2 transverse dimensions of millimeter size that are
T-dual to the 2 heterotic string size coordinates. This is depicted in
the following diagram, together with the case $k=6$, where we use
heterotic length units $l_H=1$:
\begin{scalepic}
{H: $k=4$}{I$^\prime$: $n=2$}{scal04}
\scaleitem{30}{$l_H$, $R_{5,6}$}{1}
\scaleitem{140}{$R_{1,2,3,4}=R$}{$l_I$}
\scaleitem{250}{$R^2$}{$\tilde R_{5,6}$}
\end{scalepic}
\begin{scalepic}
{H: $k=6$}{I$^\prime$: $n=6$}{scal06}
\scaleitem{30}{$l_H$}{1}
\scaleitem{140}{$R_{1,\cdots,6}=R$}{}
\scaleitem{195}{$R^{3/2}$}{$l_I$}
\scaleitem{250}{$R^2$}{$\tilde R_{1,\cdots,6}$}
\end{scalepic}

We will now show that the  low-scale type II models describe some
strongly coupled heterotic vacua and, in particular, the cases with
$k=1,2,3$ large dimensions that have not a perturbative description in
terms of type I$^\prime$ theory~\cite{ap}. In 6 dimensions, the heterotic $E_8\times E_8$
superstring compactified on $T^4$ is S-dual to type IIA compactified on
$K3$ \cite{ht}:
\be
\lambda_{6IIA}={1\over\lambda_{6H}}\qquad\qquad
l_{IIA}=\lambda_{6H}l_H\, ,
\label{het-II}
\ee
which can be obtained, for instance, by comparing eqs.(\ref{IIA}) with
(\ref{het}), using $\lambda_{6H}=\lambda_H l_H^2/\sqrt{V_{T^4}}$.
However, in contrast to the case of heterotic -- type I/I$^\prime$
duality, the compactification manifolds on the two sides are not the same
and a more detailed analysis is needed to study the precise mapping of
$T^4$ to $K3$, besides the general relations (\ref{het-II}).

This can be done through M-theory and one finds that
the different radii satisfy the following relations, in corresponding string units:
\be
{R_I\over l_{IIA}}={V_{T^4}^{1/2}\over l_H^2}\qquad\qquad
{R_1\over l_H}={V_{K3}^{1/2}\over l_{IIA}}\, '
\label{RV}
\ee
and 
\be
{R_i\over R_j}={{\tilde R}_i\over{\tilde R}_j}\qquad\qquad i,j=2,3,4\, ,
\ee
which yields ${\tilde R}_i=l_M^3/(R_jR_k)$ with $i\ne j\ne k\ne i$ and
$l_M^3=\lambda_H l_H^3$. Here $R_I$ is defined as the radius of $S^1/Z_2$ 
appearing in $K3$ when it is  ``squashed" to the shape of
$S^1/Z_2(R_I)\times T^3({\tilde R}_2,{\tilde R}_3,{\tilde R}_4)$.
This relation, together with eq.(\ref{RV}),
gives the precise mapping between $T^4$ and $K3$, which completes the
S-duality transformations (\ref{het-II}). We recall that on the type II
side, the four $K3$ directions corresponding to $R_I$ and ${\tilde R}_i$
are transverse to the 5-brane where gauge interactions are localized.

Using the above results, one can now study the possible perturbative type
II descriptions of 4d heterotic compactifications on
$T^4(R_1,\cdots,R_4)\times T^2(R_5,R_6)$ with a certain number $k$ of
large dimensions of common size $R$ and string coupling 
$\lambda_H\sim (R/l_H)^{k/2}\gg 1$. From eq.(\ref{het-II}), the type II
string tension appears as a non-perturbative threshold at energies of the
order of the $T^2$ compactification scale, $l_{II}\sim\sqrt{R_5R_6}$.
Following the steps we used in the context of heterotic -- type I duality,
after T-dualizing the radii which are smaller than the string size, one
can easily show that the $T^2$ directions must be among the $k$ large
dimensions in order to obtain a perturbative type II description. 

It follows that for $k=1$ with, say, $R_6\sim R\gg l_H$, the
type II threshold appears at an intermediate scale
$l_{II}\sim\sqrt{Rl_H}$, together with all 4 directions of $K3$, while
the second, heterotic size, direction of $T^2$ is T-dual (with respect to
$l_{II}$) to $R$: ${\tilde R}_5\equiv l_{II}^2/l_H\sim R$. Thus, one
finds a type IIB description with two large longitudinal dimensions along
the $T^2$ and string coupling of order unity, which is the example
discussed in sections {\it 2.3} and {\it 3.2}.
\begin{scalepic}
{H: $k=1$}{IIB, $\lambda\!\!\sim\!\!1$}{scal01}
\scaleitem{40}{$l_H$, $R_{1,\cdots,4}, R_5$}{1}
\scaleitem{110}{$\sqrt{R}$}{$l_{IIB}$, $K3$}
\scaleitem{180}{$R_6=R$}{$T^2({\tilde R}_5,R_6)$}
\end{scalepic}
For $k\ge 2$, the type II scale
becomes of the order of the compactification scale, $l_{II}\sim R$. For
$k=2$, all directions of $K3\times T^2$ have the type II size, while the
type II  string coupling is infinitesimally small, $\lambda_{II}\sim
l_H/R$, which is the example discussed in section {\it 3.2}. 
\begin{scalepic}
{H: $k=2$}{II,$\lambda\!\!\sim\!\!1\!/\!R$}{scal02}
\scaleitem{40}{$l_H$, $R_{1,\cdots,4}$}{1}
\scaleitem{180}{$R_{5,6}=R$}{$l_{II}$, $K3$, $T^2(R_{5,6})$}
\end{scalepic}
For $k=3$, 
$l_{II}\sim R_{5,6}\sim R$, while the four (transverse) directions of
$K3$ are extra large: $R_I\sim{\tilde R}_i\sim R^{3/2}/l_H$.
\begin{scalepic}
{H: $k=3$}{II, $\lambda\!\!\sim\!\!1$}{scal03}
\scaleitem{40}{$l_H$, $R_{2,3,4}$}{1}
\scaleitem{180}{$R_1=R_{5,6}=R$}{$l_{II}$, $T^2(R_{5,6})$}
\scaleitem{250}{$R^{3/2}$}{$K3$}
\end{scalepic}

For $k=4$, the type II dual theory provides a perturbative description
alternative to the type I$^\prime$ with $n=2$ extra large transverse
dimensions. For $k=5$, there is no perturbative type II description,
while for $k=6$, the heterotic $E_8\times E_8$ theory is described by a
weakly coupled type IIA with all scales of order $R$ apart one $K3$
direction ($R_I$) which is extra large. This is equivalent to type
I$^\prime$ with $n=1$ extra large transverse dimension.

\section{ Theoretical implications }

We will now focus on some theoretical implications of the low scale
string scenario. Unless explicitly stated otherwise, we will restrict
ourselves to the context of type I strings.

\subsection {U.V./ I.R. correspondence}

In addition to the open strings decscribing the gauge degrees of
freedom, consistency of string theory requires the presence of closed strings
associated with gravitons and different kind of moduli fields $m_a$. 

There are two types of extended objects: $D$-branes and orientifolds.
The former are hypersurfaces on which open strings end while the latter are 
hypersurfaces located at fixed points  when acting simultaneousely with a
$Z_2$ parity on the transverse  space and world-sheet coordinates.

Closed strings can be emitted by $D$-branes and orientifolds, the lowest
order diagrams being discribed by a cylinderic topology. In this way D-branes
and orientifolds appear as to lowest order classical point-like sources in the
transverse space.  For weak  type-I string coupling this can be 
described by a lagrangian of the form
 
\be
\int d^n x_\perp \; \Bigl[ {1\over g_s^2}
(\partial_{x_\perp}  m_a)^2 +  {1\over g_s} \sum_s f_s(m_a) 
\delta(x_\perp - {x_\perp}_s)\Bigr] \, ,
\label{b}
\ee
where ${x_\perp}_s$ is the location of the source $s$ ($D$-branes and
orientifolds) while $f_s(m_a)$ encodes the coupling of this source to
the moduli $m_a$. As a result while  $m_a$  have constant values in the 
four-dimensional space, their  expectation values will generically vary as a 
function of the transverse coordinates $x_\perp$ of the $n$ directions with 
 size $\sim R_\perp$ large compared to the string length $l_s$. 

Solving the classical equation of motion  for $m_a$ in (\ref{b}) leads to
contributions to the parameters (couplings) on the brane of the low energy
effective action  given by a sum of Green's functions of the form~\cite{ab}:
\be
  {1\over V_\perp}\  \sum_{\vert {p}_\perp\vert < M_{s} }\
{1\over p_\perp^2}\  F({\vec p_\perp})\, ,
\label{tadpole}
\ee 
where $V_\perp={R_\perp}^{d_\perp}$ is the volume of the
transverse space, ${\vec p}_\perp =(m_1/{R_\perp}$$\cdots$
${m_{d_\perp}/{R_\perp}})$ is the transverse momentum exchanged  by
the massless closed string, $F({\vec p_\perp})$ are the
Fourier-transformed to momentum space of derivatives of $f_s(m_a)$. 
An explicit
expression can be given in the simple case of toroidal
compactification with vanishing antisymmetric tensor, where the global tadpole cancelation fixes  the number of D-branes to be 32:
\be
F({\vec p_\perp})\sim\left( 32\prod_{i=1}^{d_\perp}{1+(-)^{m_i}\over 2}
-2\sum_{a=1}^{16}{\rm cos}({\vec p_\perp}{\vec x_a})\right)\, ,
\label{tadpole2}
\ee
where  ${\vec p}_\perp =(m_1/{R_\perp}$$\cdots$
${m_{d_\perp}/{R_\perp}})$, the orientifolds are located at the corners of the cell 
$[0, \pi R_\perp]^{d_\perp}$ and are responsible for the first term in 
(\ref{tadpole2}), and $\pm{\vec x_a}$ are the transverse
positions of the 32 D-branes (corresponding to Wilson lines of the
T-dual picture) responsible of the second term.

In a compact space where flux lines can not escape to infinity, the Gauss-law 
implies that the total charge, thus global tadpoles, should vanish $F(0)=0$ 
while local tadpoles may not vanish $F({\vec p_\perp}) \neq 0$ for $\vec p
\neq 0$. In that case, obtained for generic positions of the D-branes, the 
tadpole contribution (\ref{tadpole}) leads to the following behavior in the
large radius limit  for the moduli $m_a$:
\be
 m_a ({x_\perp}_s) \sim \cases{ O(R_\perp M_s)\ &for \ \ \ $d_\perp=1$\cr
O(\ln R_\perp M_s)\ \ \ &for \ \ \ $d_\perp=2$\cr
O(1) \ &for\ \ \ \ $d_\perp>2$\cr}\quad ,
\ee
which is dictated by the large-distance behavior of the two-point Green 
function in the $d_\perp$-dimensional transverse space.

There are some important implications of these results:

\begin{itemize} 

\item The tree-level exchange diagram of a closed string can also be seen
as  one-loop exchange of open strings. While from the former point of
view, a long cylinder represents an infrared limit where one computes
the effect of exchanging light closed strings at long distances, in the
second point of view the same diagram is conformally mapped to an
annulus describing the one-loop running in the ultraviolet limit of
very heavy open strings streching between the two  boundaries of the
cylinder. Thus, from the brane gauge theory point of view, there are
ultraviolet effects that are not cut-off by the string scale $M_s$ but
instead by the winding mode scale $R_\perp M_s^2$.

\item In the case of one large dimension $d_\perp=1$, the corrections are 
linear in $R_\perp$. Such correction appears for instance for the dilaton field
which sits in front of gauge kinetic terms, that drive the
theory rapidly to a strong coupling singularity and, thus, forbid the
size of the transverse space to become much larger than the string length.
It is possible to avoid such large corrections if the
tadpoles cancel locally. This happens when
D-branes are equally distributed at the two fixed points of the
orientifold. 

\item The case $d_\perp=2$ is particularly attractive because it
allows the effective couplings of the brane theory to depend
logarithmically on the size of the transverse space, or equivalently
on $M_P$, exactly as in the case of softly broken supersymmetry at
$M_s$.  Both higher derivative and higher string loop corrections to
the bulk supergravity lagrangian are expected to be small for slowly
(logarithmically) varying moduli. The {\it classical} equations of motion of 
the effective 2d
supergravity in the transverse space are analogous to the
renormalization group equations used to  resum  large corrections to
the effective field  theory parameters with appropriate boundary
conditions.

\end{itemize}

It turns out that low-scale type II theories with infinitesimal string
coupling share many common properties with type I$^\prime$ when
$d_\perp=2$~\cite{ap}. In fact, the limit of vanishing coupling does not
exist due to subtleties related to the singular character of the
compactification manifold and to the non perturbative origin of gauge
symmetries. In general, there are corrections depending logarithmically
on the string coupling, similarly to the case of type I$^\prime$ strings
with 2 transverse dimensions.

\subsection {Unification}

One of the main succes of low-energy supersymmetry is that the three
gauge  couplings of the Standard Model, when extrapolated at high
energies assuming the particle content of its $N=1$ minimal
supersymmetric extension (MSSM), meet at an energy scale $M_{\rm
GUT}\simeq 2\times 10^{16}$ GeV. This running is described at the the
one-loop level by: 
\be {1\over g^2_a(\mu)}={1\over g^2}+{b_a\over
4\pi} \ln{M_{\rm GUT}^2\over\mu^2}\, , \ee 
where $\mu$ is the energy
scale and $a$ denotes the 3 gauge group factors of the Standard Model
$SU(3)\times SU(2)\times U(1)$. Note that even in the absence of any
$GUT$ group, if one requires keeping unification of all gauge couplings
then the string relations we discussed in section {\it 3}
suggest that the gauge theories arise from the same kind of branes.

Decreasing the string scale below energies of order $M_{GUT}$ is
expected to cut-off the runing of the couplings before they meet and
thus spoils the unification. Is there a way to reconcile the apparent
unification with a low string scale?

One possibility is to use power-law running that may
accelerate unification in an energy region where the theory becomes
higher dimensional~\cite{ddg}. Within the effective field theory, the
summation over the KK modes above the compactification scale and below
some energy scale
$E\gg R^{-1}$ yields:
\be
{1\over g^2_a(E)}={1\over g^2_a(R^{-1})}-{b_a^{SM}\over 2\pi}\ln(ER)-
{b_a^{KK}\over 2\pi}\left\{ 2\left( ER-1\right)-\ln(ER)\right\}\, ,
\label{powerev}
\ee
where we considered one extra (longitudinal)
dimension. The first logarithmic term corresponds to the usual 4d
running controlled by the Standard Model beta-functions $b_a^{SM}$, while
the next term is the contribution of the KK tower dominated by the
power-like dependence $(ER)$ associated to the effective multiplicity
of KK modes and controlled by the corresponding beta-functions
$b_a^{KK}$.

Supersymmetric theories in higher dimensions have at least$N=2$
extended supersymmetry thus   the KK excitations form supermultiplets of 
$N=2$. There are two kinds of such supermultiplets, the vector
multiplets containing spin-1 field, a Dirac fermion and 2 real scalars
in the adjoint representation and  hypermultiplets  containing 
an $N=1$ chiral multiplet and its mirror. As the gauge degrees of freedom
are to be identified with bulk fields, their KK excitations will be part of 
$N=2$ vector multiplets. The higgs and matter fields, quarks and leptons, can 
on the other hand be chosen to be either localized without KK 
excitations or instead identified with bulk states with KK excitations forming 
$N=2$ hypermultiplets representations. Analysis of unification with 
the corresponding coefficients  
$b_a^{KK}$  has been performed in \cite{poweruni}.

There are two remarks to be made on this approach: (i) the result
is very sensitive (power-like) to the initial conditions and thus to
string threshold corrections, in contrast to the usual unification based
on logarithmic evolution, (ii) only the case of one extra-dimension
appears to lead to power-like corrections in type I models.

In fact the one-loop corrected gauge couplings in $N=1$ orientifolds are
given by the following expression~\cite{abd}:
\be
{1\over g^2_a(\mu)}={1\over g^2}+s_a m+
{b_a\over 4\pi}\ln{M_I^2\over\mu^2}-\sum_{i=1}^3{b_{a,i}^{N=2}\over 4\pi}
\left\{\ln T_i +f(U_i)\right\}\, ,
\label{thresholds}
\ee
where the first two terms in the r.h.s. correspond to the tree-level
(disk) contribution and the remaining ones are the one-loop (genus-1)
corrections. Here, we assumed that all gauge group factors correspond to
the same type of D-branes, so that gauge couplings are the same to lowest
order (given by $g$). $m$ denotes a combination of the twisted moduli,
whose VEVs blow-up the orbifold singularities and allow the transition to
smooth (Calabi-Yau) manifolds. However, in all known examples, these VEVs
are fixed to $m=0$ from the vanishing of the D-terms of anomalous
$U(1)$'s.

As expected, the one-loop corrections contain an infrared divergence,
regulated by the low-energy scale $\mu$, that produces the usual 4d
running controlled by the $N=1$ beta-functions $b_a$. The last sum
displays the string threshold corrections that receive contributions only
from $N=2$ sectors, controlled by the corresponding $N=2$
beta-functions $b_{a,i}^{N=2}$. They depend on the geometric moduli
$T_i$ and $U_i$, parameterizing the size and complex structure of the
three internal compactification planes. In the simplest case of a
rectangular torus of radii $R_1$ and $R_2$, $T=R_1R_2/l_s^2$ and
$U=R_1/R_2$. The function $f(U)=\ln\left({\rm Re}U|\eta(iU)|^4\right)$
with $\eta$ the Dedekind-eta function; for large $U$, $f(U)$ grows
linearly with $U$.  Thus, from expression (\ref{thresholds}), it follows
that when $R_1\sim R_2$, there are logarithmic corrections (as explained
 for transverse directions to the brane for the previous section)
$\sim\ln(R_1/l_s)$, while when $R_1>R_2$, the corrections grow linearly
as $R_1/R_2$. Note that in both cases, the corrections are proportional
to the $N=2$ $\beta$-functions and there no power law corrections in the case 
of more than one large compact dimensions.

Obviously, unification based on
logarithmic evolution requires the two (transverse) radii to be much
larger than the string length, while power-low unification can happen
either when there is one longitudinal dimension a bit larger than the
string scale  ($R_1/R_2\sim R_\parallel/l_s$ keeping $g_s<1$), or
when one transverse direction is bigger than the rest of the bulk.

The most advantageous possibility is to obtain large logarithmic thresholds
depending on two large dimensions transverse to the brane ($d_\perp=2$).
One hopes that such logarithmic
corrections may restore the ``old" unification picture with a GUT
scale given by the winding scale, which for millimeter-size dimensions
has the correct order of magnitude~\cite{cb,ab,admr}. In this way, the
running due to a large desert in energies is replaced by an effective
running due to a ``large desert" in transverse distances from our
world-brane. However, the logarithmic contributions are model
dependent~\cite{abd} and at present there is no compelling explicit
realization of this idea.

\subsection{Supersymmetry breaking and scales hierarchy}

When decreasing the string scale, the question of hierarchy of scales i.e. of
why the Planck mass is much bigger than the weak scale, is translated
into the question of why there are transverse dimensions much larger
than the string scale, or why the string coupling is very
small. For instance for a string scale in the TeV range, 
 From eq.(\ref{treei}) in type I/I$^\prime$ strings, the required
hierarchy $R_\perp/l_I$ varies from $10^{15}$ to $10^5$, when the number
of extra dimensions in the bulk varies from $n=2$ to $n=6$, respectively,
while in type II strings with no large dimensions, the required value of
the coupling $\lambda_{II}$ is $10^{-14}$.

There are two issues that one needs to address:

\begin{itemize}

\item We have seen in section 4.1 that although the string 
scale is very low, there might be large quantum corrections that arise, 
dependending on the size of the large dimensions transverse to the brane.
This is as if the UV
cutoff of the effective field theory on the brane is not the
string scale but the winding scale $R_\perp M_I^2$, dual to the large
transverse dimensions and which can be much larger than the string scale.
In particular such correction could spoil the nullification of gauge hierarchy 
that remain the main theoretical motivation of TeV scale strings.

\item Another important issue is to understand the dynamical question on 
the origin of the hierarchy.
 
\end {itemize}

TeV scale strings offer
a solution to the technical (at least) aspect of gauge hierarchy without
the need of supersymmetry, provided there is no effective propagation of
bulk fields in a single transverse dimension, or else closed string
tadpoles should cancel locally. The case of $d_\perp =2$ leads to 
a logarithmic dependence of the effective potential
 on $R_\perp / l_s$ which allows the possible radiative generation  of the 
hierarechy between $R_\perp$ and $l_s$ as for no-scale models. Moreover, it
is interesting to notice that the ultraviolet behavior of the theory is very 
similar with the one with soft supersymmetry breaking at $M_s \sim TeV$.  It
is then natural to ask the question whether there is any motivation leftover
for supersymmetry or not. This bring  us to the problems of the stability of
the new hierarchy and  of the cosmological constant~\cite{aadd}.

In fact, in a non-supersymmetric string theory, the bulk energy density
behaves generically as $\Lambda_{\rm bulk}\sim M_s^{4+n}$, where $n$ is
the number of transverse dimensions much larger than the string length.
In the type I/I$^\prime$ context, this induces a cosmological constant on
our world-brane which is enhanced by the volume of the transverse space
$V_\perp\sim R_\perp^n$. When expressed in terms of the 4d parameters
using the type I/I$^\prime$ mass-relation (\ref{treei}), it is translated
to a quadratically dependent contribution on the Planck mass:
\be
\Lambda_{\rm brane}\sim M_I^{4+n}R_\perp^n\sim M_I^2 M_P^2\, ,
\label{lambda}
\ee
where we used $s=I$. This contribution is in fact the analogue of the
quadratic divergent term Str${\cal M}^2$ in softly broken supersymmetric
theories, with $M_I$ playing the role of the supersymmetry breaking
scale. 

The brane energy density (\ref{lambda}) is far above the (low) string
scale $M_I$ and in general destabilizes the hierarchy that one tries to
enforce. One way out is to resort to special models with broken
supersymmetry and vanishing or exponentially small cosmological
constant~\cite{ks}. Alternatively, one could conceive a different
scenario, with supersymmetry broken primordially on our world-brane
maximally, i.e. at the string scale which is of order of a few TeV. In
this case the brane cosmological constant would be, by construction,
${\cal O}(M_I^4)$, while the bulk would only be affected by
gravitationally suppressed radiative corrections and thus would be almost
supersymmetric~\cite{aadd,ads}. In particular, one would expect the
gravitino and other soft masses in the bulk to be  extremely small
$O(M_I^2/M_P)$. In this case, the cosmological constant induced in the
bulk would be
\be
\Lambda_{\rm bulk}\sim M_I^4/R_\perp^n\sim M_I^{6+n}/M_P^2\, ,
\label{lambdasmall}
\ee
i.e. of order (10 MeV)$^6$ for $n=2$ and $M_I\simeq 1$ TeV.
The scenario of brane supersymmetry breaking is also required in models
with a string scale at intermediate energies $\sim 10^{11}$ GeV (or
lower), discussed in the beginning of section {\it 3}. It can occur for instance on a
brane distant from our world and is then mediated to us by gravitational
(or gauge) interactions.

In the absence of gravity, brane supersymmetry breaking can
occur in a non-BPS system of rotating or intersecting D-branes. Since
brane rotations correspond to turning on background magnetic fields, they
can be easily generalized in the presence of gravity, in the context of
type I string theory~\cite{ba}. Stable non-BPS configurations of intersecting 
branes
have been studied more recently, while their implementation in
type I theory was achieved in Ref.~\cite{ads}.

The simplest examples are based on orientifold projections of type IIB,
in which some of the orientifold 5-planes have opposite charge, requiring
an open string sector living on anti-D5 branes in order to cancel the RR
(Ramond-Ramond) charge. As a result, supersymmetry is broken on the
intersection of D9 and anti-D5 branes that coincides with the world
volume of the latter. The simplest construction of this type is a 
$T^4/Z_2$ orientifold with a flip of the $\Omega$-projection (world-sheet
parity) in the twisted orbifold sector. It turns out that several
orientifold models, where tadpole conditions do not admit naive
supersymmetric solutions, can be defined by introducing
non-supersymmetric open sector containing anti-D-branes. A typical
example of this type is the ordinary $Z_2\times Z_2$ orientifold with
discrete torsion. 

The resulting models are chiral, anomaly-free, with vanishing RR tadpoles
and no tachyons in their spectrum~\cite{ads}. Supersymmetry is broken at
the string scale on a collection of anti-D5 branes while, to lowest
order, the closed string bulk and the other branes are supersymmetric. In
higher orders, supersymmetry breaking is of course mediated to the
remaining sectors, but is suppressed by the size of the transverse space
or by the distance from the brane where supersymmetry breaking primarily
occurred. The models contain in general uncancelled NS (Neveu-Schwarz)
tadpoles reflecting the existence of a tree-level potential for the NS
moduli, which is localized on the (non-supersymmetric) world volume of
the anti-D5 branes.

As a result, this scenario implies the absence of supersymmetry on our
world-brane but its presence in the bulk, a millimeter away! The bulk
supergravity is needed to guarantee the stability of gauge hierarchy
against large gravitational quantum radiative corrections.\\

\subsection {Electroweak symmetry breaking in TeV-scale strings}

The existence of non-supersymmetric type I string vacua allows us to address
the question of gauge symmetry breaking.  From the effective field theory point of
 view, one expects quadratic divergences in one-loop 
contribution to the masses of scalar fields. It is then imoportant to 
address the following questions: (i) which scale plays the role of the 
Ultraviolet cut-off (ii) could these one-loop corrections be used to 
to generate radiatively the electroweak symmetry breaking~\footnote{For an earlier attempt to
generate a non-trivial
minimum of the potential, see Ref.~\cite{gg}.}, and 
explain the mild hierarchy between the weak and a  string scale at a few TeVs.

A simple framework to address such 
issues is non-supersymmetric tachyon-free $Z_2$ orientifold 
of type IIB superstring compactified to four dimensions on
$T^4/Z_2\times T^2$~\cite{ads}. Cancellation of Ramond-Ramond charges
requires the presence of 32 D9 and 32 anti-D5 (D$\bar5$)
branes. The bulk (closed strings) 
as well as the D9 branes are
$N=2$ supersymmetric while supersymmetry is broken on the world-volume of the
D$\bar5$'s. It is possible \cite{abqhiggs} to compute the effective potential involving the
scalars of the D$\bar5$ branes, namely in this simple example 
the adjoints and bifundamentals
of the $USp(16)\times USp(16)$ gauge group. The
resulting potential has a non-trivial minimum which fixes the 
VEV of the Wilson line or, equivalently, the
distance between the branes in the $T$-dual picture. Although the
obtained VEV is of the order of the string scale, the potential
provides a negative squared-mass term when expanded around the
origin:

In the limit where the radii of the transverse space are large, 
$R_\perp \to\infty$ and for arbitrary longitudinal radius $R_\parallel$,
the result is:
\be
\label{mu2R}
\mu^2(R_\parallel)=-\varepsilon^2(R_\parallel)\, g^2\, M_s^2
\ee
with
\be
\varepsilon^2(R_\parallel) ={1\over 2\pi^2}\int_0^\infty \frac{dl}{\left(2\,
l\right)^{5/2}} 
{\theta_2^4\over 4\eta^{12}}\left(il+{1\over 2}\right) R_\parallel^3
\sum_n n^2 e^{-2\pi n^2R_\parallel^2l}\ .
\label{epsilon2R}
\ee

For the 
asymptotic value $R_\parallel\to 0$~\footnote{This limit corresponds, 
upon T-duality,
to a large transverse dimension of radius $1/R_\perp$.},
$\varepsilon(0)\simeq 0.14$, and the effective cutoff for the mass
term at the origin is $M_s$, as can be seen from Eq.~(\ref{mu2R}). At
large $R_\parallel$, $\mu^2(R_\parallel)$ falls off as $1/R_\parallel^2$, which is the effective
cutoff in the limit $R_\parallel\to\infty$, in agreement with field theory
results in the presence of a compactified extra
dimension~\cite{SS}~\footnote{Actually this effect is at the origin of thermal
squared masses, $\sim T^2$, in four-dimensional field theory at finite
temperature, $T$, where the time coordinate is compactified on a circle
of inverse radius $1/R_\parallel\equiv T$ and the Boltzmann suppression factor
generates an effective cutoff at momenta $p \sim T$.}. In fact, in the
limit $R_\parallel\to\infty$ an analytic approximation to $\varepsilon(R)$ 
gives:
\be
\varepsilon(R_\parallel)\simeq \frac{\varepsilon_\infty}{M_s\, R_\parallel}\, ,
\qquad\qquad
\varepsilon_\infty^2=\frac{3\, \zeta(5)}{4\, \pi^4}\simeq 0.008\, .
\label{largeR}
\ee

While the mass term (\ref{mu2R}) was computed for the Wilson line it 
also applies, by gauge invariance, to the charged massless fields
which belong to the same representation. By 
orbifolding the previous example, the Wilson line is projected 
away from the spectrum and we are left with  the 
charged massless fields with
quartic tree-level terms and  one-loop negative
squared masses.
By identifying them with the Higgs field we can achieve radiative
electroweak symmetry breaking, and obtain the mild hierarchy
between the weak and string scales in terms of a loop factor. 
More precisely, in the minimal case where there is only one such Higgs
doublet $h$, the scalar potential would
be:
\be
V=\lambda (h^\dagger h)^2 + \mu^2 (h^\dagger h)\, ,
\label{potencialh}
\ee
where $\lambda$ arises at tree-level and is given by an appropriate truncation
of a supersymmetric theory. This property remains valid in any model where
the higgs field comes from an open string with both ends fixed on the same type
of D-branes (untwisted state).
Within the minimal spectrum of the Standard Model,
$\lambda=(g_2^2+g'^2)/8$, with $g_2$ and $g'$ the $SU(2)$ and $U(1)_Y$ gauge
couplings, as in the MSSM. On the other hand,
$\mu^2$ is generated at one loop and can be estimated by
Eqs.~(\ref{mu2R}) and (\ref{epsilon2R}).

The potential (\ref{potencialh}) has a minimum at $\langle h\rangle
=(0,v/\sqrt{2})$, where $v$ is the VEV of the neutral component of the $h$
doublet, fixed by $v^2=-\mu^2/\lambda$. Using the relation of $v$ with the $Z$
gauge boson mass, $M_Z^2=(g_2^2+g'^2)v^2/4$, and the fact that the quartic
Higgs interaction is provided by the gauge couplings as in supersymmetric
theories, one obtains for the Higgs mass a prediction which is the
MSSM value for $\tan\beta\to\infty$ and $m_A\to\infty$:
\be
\label{masa}
M_h=M_Z\ .
\ee
Furthermore, one can compute $M_h$ in terms of the string scale
$M_s$, as $M_h^2=-2\mu ^2=2\varepsilon^2 g^2M_s^2$, or equivalently
\begin{equation}
M_s=\frac{M_h}{\sqrt{2}\, g\varepsilon}
\label{final}
\end{equation}

The determination of the precise value of the string scale suffers from two
ambiguities. The first is the value of the gauge coupling $g$ at $M_s$,
which depends on the details of the model. A second ambiguity concerns the
numerical coefficient
$\varepsilon$ which is in general model dependent.
Varying $R$
from 0 to 5, that covers the whole range of values for a transverse dimension
$1<1/R_\perp <\infty$, as well as a reasonable range for a longitudinal dimension
$1<R_\parallel \simlt 5$, one obtains $M_s\simeq 1-5$ TeV. 
In the $R_\parallel\gg 1$ (large longitudinal dimension) region our
theory is effectively cutoff by $1/R_\parallel$ and the Higgs mass is then
related to it by, 
\begin{equation}
\frac{1}{R_\parallel}=\frac{M_h}{\sqrt{2}\, g\,\varepsilon_\infty}\, .
\label{finalR}
\end{equation}
Using now the value for $\varepsilon_\infty$ in the present model,
Eq.~(\ref{largeR}), we find $1/R_\parallel\simgt 1$ TeV.

The tree level Higgs mass has been shown to receive important
radiative corrections from the top-quark sector. For present experimental
values of the top-quark mass, the Higgs mass in Eqs.~(\ref{masa}) and
(\ref{final}) is raised to values around 120 GeV~\cite{higgs}.
In addition there might be large string threshold corrections.
To illustrate these issue consider  the relevant part of the world brane action in the
string frame in the simplest case:
\bea
{\cal L}_{\rm brane}&=&e^{-\phi}\left\{\omega^2|DH|^2+{1+\tan^2\theta_W\over 8}
\omega^4(H^\dagger H)^2 +{1\over 4}(F_{SU(2)}^2+\cot^2\theta_W F_Y^2)
\right\} \cr &-&\varepsilon^2M_s^2\omega^4|H|^2\, ,
\label{Lbrane}
\eea
where $\phi$ is the string dilaton, $\omega$ the scale factor of the
four-dimensional (world brane) metric, $H$ the Higgs scalar (in the string
frame) and $D$ the gauge covariant derivative. The weak angle at the string
scale $\theta_W$ must be correctly determined in the string model.
Notice that the last term has no $e^\phi$ dependence since it corresponds to a
one loop correction. The bulk fields $\phi$ and $\omega$ are evaluated in the
transverse coordinates at the position of the brane. The physical couplings
$g_2$, $\lambda$ and the mass $\mu^2$ are given by
\be
g_2=e^{\phi/2}\ ,\qquad
\lambda={1+\tan^2\theta_W\over 8}\, e^\phi\ ,\qquad 
\mu^2=-\varepsilon^2e^\phi\omega^2M_s^2\, ,
\label{param}
\ee
while Eq.~(\ref{masa}) remains unchanged and the relation (\ref{final}) becomes
\be
\label{finalc}
M_s=\frac{M_h}{\sqrt{2}\, \varepsilon\, e^{\phi/2}\omega}\, .
\ee
The lowest order result (\ref{final}) corresponds to the (bare) value $\omega=1$.

As we discussed in section 4.1, when the bulk fields $\phi$ and $\omega$ propagate in
two large transverse dimensions, they acquire a logarithmic dependence on
these coordinates due to distant sources. Since the value of $\phi$ at the
position of the world brane is fixed by the value of the gauge coupling in
Eq.~(\ref{param}), the relation (\ref{masa}) for the Higgs mass is not
affected, while Eq.~(\ref{finalc}) for the string scale is corrected by a
renormalization of $\omega$ which takes the generic form:
\be
\omega=1+b_\omega g_2^2\ln (R_\perp M_s)\, ,
\label{omega}
\ee
where $b_\omega$ is a numerical coefficient. This correction is similar to a
usual renormalization factor in field theory, which here is due to an infrared
running in the transverse space. Depending on the sign of $b_\omega$, it can
enhance ($b_\omega<0$) or decrease ($b_\omega>0$) the value of the string scale
by the factor $1/\omega$. This effect can be important since the involved
logarithm is large, varying between 7 and 35, for $R_\perp$ between 1 fm and 1
mm.

\section{ Scenario for studies of experimental constraints }\label{sec:exp}

In order to pursue further, we need to provide the quantum numbers and 
couplings of the relevant light states. In the scenario we consider:
\begin{itemize}

\item Gravitons~\footnote{ Along with gravitons, string models predict the
presence of other very weakly coupled states as gravitinos, dilatons,
moduli, Ramond-Ramond fields....These might alter the bounds obtained in 
Section~\ref{subsec:miss}.} which describe fluctuations of the metric 
propagate in the whole 10- or 11-dimensional space. 

\item In all generality,  gauge bosons propagate on a
$(3+d_\parallel)$-brane, with  $d_\parallel=0,...,6$. However, as we have seen
in the previous sections, a freedom of  choice for the values of the
string and compactification scales requires that gravity and gauge
degrees of freedom live in spaces with different
dimensionalities. This means that $d_{\parallel max} =5$ or 6 for 10-
or 11-dimensional theories, respectively. The value of $d_\parallel$ represents
the number of dimensions felt by KK excitations of gauge bosons.

To simplify the discussion, we will mainly consider the case $d_\parallel=1$ where some
of the gauge fields arise from a 4-brane. Since the  couplings  of the
corresponding gauge groups are reduced by  the size of the large dimension
$R_\parallel M_s$ compared to the others, if  $SU(3)$ has KK modes all three
group factors must have. Otherwise it is difficult to reconcile the
suppression of the strong coupling at the string scale with the observed
reverse situation. As a result, there are 5 distinct cases \cite{AAB} 
that we denote
$(l,l,l)$, $(t,l,l)$, $(t,l,t)$, $(t,t,l)$ and $(t,t,t)$, where the three
positions in the brackets correspond to the 3 gauge group factors of the
standard model $SU(3)_c\times SU(2)_w\times U(1)_Y$ and those with $l$ feel
the extra-dimension, while those with $t$ (transverse) do not.

\item The matter fermions, quarks and leptons, are localized on the intersection of a
3-brane with the $(3+d_\parallel)$-brane and have no KK excitations along 
the $d_\parallel$ directions.  
Their coupling to KK modes of gauge bosons are given in Eq.~\ref{coupling}.
This is the main assumption in our analysis and limits derived in the next
subsection depend on it. In a more general study it could be relaxed by
assuming that only part of the fermions are localized. However, if all
states are propagating in the bulk, then the KK excitations are stable and a
discussion of the cosmology will be necessary in order to explain why they
have not been seen as isotopes.

Let's denote generically the localized states as $T$ while the bulk states with KK momentum $n/R$ by $U_n$, thus the only trilinear allowed couplings are
$g_nTT U_n$ and $g U_n U_m U_{n+m}$ where $g_n$ is given by Eq. (\ref{coupling}). Hence because matter fields are localized, 
their interactions do not preserve
the 
momenta in the extra-dimension and single KK excitations can be produced.
This means for example that QCD processes $q{\bar q} \rightarrow G^{(n)}$
with $q$ representing quarks and $G^{(n)}$ massive KK excitations 
of gluons are allowed. In contrast, processes such as $G G \rightarrow
G^{(n)}$
are forbidden as gauge boson interactions conserve the internal momenta.

The possible localization of the Higgs scalars will be discussed in section 6.3, as well as the possible existence of supersymmetric partners although they 
do not lead to important modifications for most of the obtained bounds.

\end{itemize}

\section{Extra-dimensions along the world brane: KK excitations of gauge
bosons}

The experimental signatures of extra-dimensions are of two types \cite{ABQ,ABQ2,AAB}:
\begin{itemize}

\item Observation of resonances due to KK excitations. This needs a collider
energy $\sqrt{s} \simgt 1/R_\parallel$ at LHC. 

\item Virtual exchange of the KK excitations which lead to measurable
deviations in cross-sections compared to the standard model prediction.

\end{itemize}
The necessary data needed to evaluate the size of 
these contributions are: the coupling constants given in (\ref{coupling}), 
the KK masses already given by (\ref{KKdef}), and the associated widths.
 The latter are given by decay rates into standard model fermions $f$:
\begin{equation}
\Gamma\left(X_n\rightarrow f\bar{f}\right)
=g^2_\alpha \frac{M_{\vec n}}{12\pi}C_f(v_f^2+a_f^2)
\label{Gammaf}
\end{equation}
and, in the case of supersymmetric brane there is an additional
contribution from decays  into the scalar superpartners
\begin{equation}
\label{Gammasf}
\Gamma\left(X_n\rightarrow \widetilde{f}_{(R,L)} 
\widetilde{\bar{f}}_{(R, L)}\right)
=g^2_\alpha\frac{M_{\vec n}}{48\pi}C_f(v_f\pm a_f)^2\, ,
\end{equation}
with $C_f = 1$ (3) for colour singlets (triplets) and $v_f, a_f$ stand for the 
standard model vector and axial couplings.\footnote{Below we will present limits for the case where the standard model particles are the only 
accessible  final states. The effect of superpatners is to enlarge the 
widths of KK excitations  by a factor 3/2.} These widths 
determine the size of corresponding resonance signals and will be important only when discussing on-shell production of KK excitations.

In the studies of virtual effects, our strategy for extracting
exclusion bounds will depend on the total number of analysed events.
If it is small then we will consider out of reach compactification
scales  which do not lead to  prediction of at least 3 new events. In
the  case of large number of events, one estimates the  deviation  from the background fluctuation ($\sim \sqrt{N_T^{\rm SM}(s)}$)  by computing the  ratio \cite{ABQ,ABQ2,AAB}
\be
\Delta_{T}=\left|{N_T(s)-N_T^{\rm SM}(s) \over \sqrt{N_T^{\rm SM}(s)}}
\right|
\label{deltaT}
\ee
where $N_T(s)$ is the total number of events while $N_T^{\rm SM}(s)$ is 
the corresponding quantity expected from the standard model. These numbers are computed using the formula:

\be
N_T =  \sigma  A \int {\cal L} dt
\label{nt}
\ee where $ \sigma$ is the relevant cross-section, $\int {\cal L} dt$
is the integrated luminosity while $A$ is a suppresion factor taking
into account the corresponding efficiency times acceptance factors.

In the next two subsections we derive limits for the case  $(l,l,l)$
where all the gauge factors feel the large extra-dimension.  We will 
return later to the other possibilities.

\subsection {Production at $e^+e^-$ colliders}

Unless the machine energy happens to be very close to their mass  KK excitation resonances will not be observed and the main expected effect will be modification of cross sections  for the $e^+e^-\rightarrow \mu^+\mu^-$ process through exchange of virtual KK excitations of the photon and $Z$ boson. Assuming unpolarized 
electron-positron pairs $e^+e^-$, the total cross section for the annihilation 
into lepton pairs  $l^+l^-$ is given by:

\begin{equation}
\sigma_T^0(s)={s\over 12 \pi} \sum_{\alpha ,\beta=\gamma, Z, KK}g^2_{\alpha}
({\sqrt s}) g^2_{\beta} ({\sqrt s}) {(v^{\alpha}_e v^{\beta}_e+
a^{\alpha}_e a^{\beta}_e)(v^{\alpha}_l v^{\beta}_l + a^{\alpha}_l
a^{\beta}_l) \over (s -m^2_{\alpha} + i\Gamma{_\alpha}
m_{\alpha})(s-m^2_{\beta} - i\Gamma_{\beta} m_{\beta})} \ ,
\label{eesig}
\end{equation}
 with $\sqrt s$ the centre--of--mass energy and the labels $\alpha ,\beta$ standing for the different neutral vector
bosons $\gamma$, $Z$, and their KK excitations.

For energies below the mass of the first KK excitation, the main 
signature will be the observation of a {\it deficit} of events due mainly to 
 interference terms. 
A precision estimate of this deficit requires inclusion of radiative corrections, and
in particular the bremsstrahlung effects on the initial electron and
positron~\cite{Dlr}. These are described by the convolution of (\ref{eesig}) 
with radiator functions, which describe the probability of having a
fractional energy loss, $x$, due to the initial--state radiation:
\begin{equation}
\sigma_T(s) =\int^{x_{\rm max}}_0 dx \sigma_T^0(s') r_T(x) ,\quad s'=s(1-x)
\label{eesigr}
\end{equation}
In the above equation, $x_{\rm max}$ represents an experimental cut off for
the energy of emitted soft photons in bremsstrahlung processes. The
radiator function is given by~\cite{Dlr}:

\begin{equation}
\label{radiator}
r_T(x) =(1 + X) y x^{y -1}+ H_T(x)\ ,
\end{equation}
with:
\bea
X &=& {e^2(\sqrt{s}) \over 4\pi^2} \left[{\pi^2\over 3}- {1\over
2}+{3\over 2}\left(\log{{s\over m_e^2}}-1\right)\right]\nonumber \\ 
y &=&{2e^2(\sqrt{s}) \over 4\pi^2} \left(\log{{s\over m_e^2}}-1\right)
\nonumber \\ 
H_T &=& {e^2(\sqrt{s}) \over 4\pi^2} \left[{1+(1-x)^2\over x}\left(\log{{s\over
m_e^2}}-1\right)\right]-{y\over x}\ ,
\eea
where $m_e$ is the electron mass.

For an estimate we use ${\sqrt s}=189$ GeV for LEPII, 500 GeV for NLC-500 and 
1 TeV for NLC-1000,  together with the numerical values for the experimental 
cuts:
\begin{eqnarray}
x_{\rm max}({\rm LEPII})&=& 0.77,\nonumber\\
x_{\rm max}({\rm NLC\rm{-}500})&=& 0.967, \nonumber\\
x_{\rm max}({\rm NLC\rm{-}1000})&=&0.992,
\end{eqnarray}
coming from imposing the cut $s'\ge M_Z^2$.

In the case of one extra dimension $d_\parallel =1$, the  sum over KK
modes in (\ref{eesig}) is dominated by the lowest modes and  converges
rapidly.  This implies: (i) the results do not depend on the value
of the string scale $M_s$, (ii) it is possible to use equal couplings
for all $n>0$ modes.

\begin{figure}[htb]
\centering
\epsfxsize=3.5in
\hspace*{0in}
\epsffile{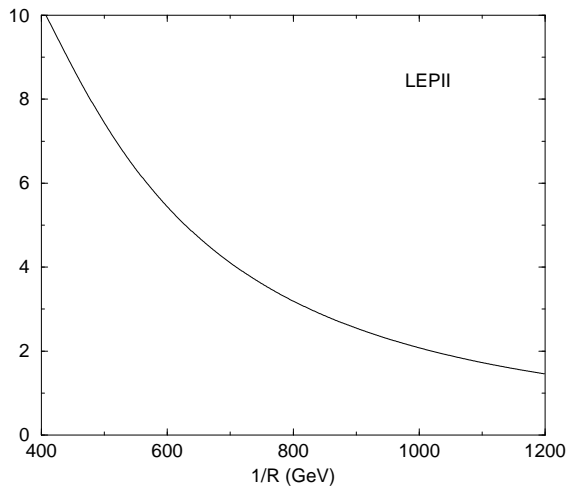}
\caption{\it Ratio $\left|{N_T(s)-N_T^{\rm SM}(s) \over \sqrt{N_T^{\rm SM}(s)}}
\right|$ from the total cross section at LEPII. We assumed a luminosity 
times efficiency of 200 $pb^{-1}$.}
\label{fig:fig1}
\end{figure}

\begin{figure}[htb]
\centering
\epsfxsize=3.5in
\hspace*{0in}
\epsffile{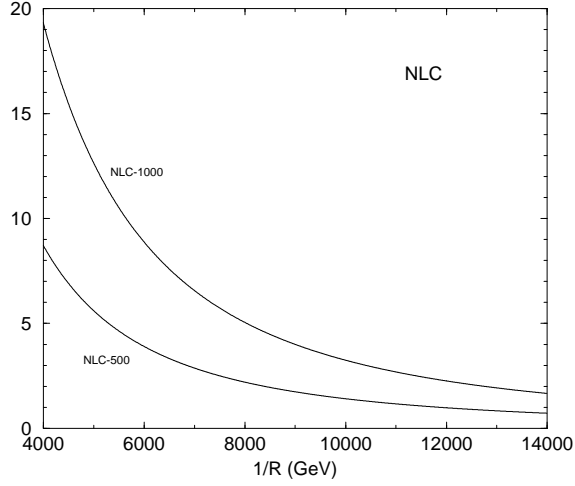}
\caption{\it Ratio $\left|{N_T(s)-N_T^{\rm SM}(s) \over \sqrt{N_T^{\rm SM}(s)}}
\right|$ from the total cross section at NLC-500 and NLC-1000. 
We assumed a luminosity times efficiency of 75 $fb^{-1}$ and 
200 $fb^{-1}$, respectively.}
\label{fig:fig2}
\end{figure}

Fig.~\ref{fig:fig1} shows the ratio $\Delta_{T}$  for LEPII while 
Fig.~\ref{fig:fig2} shows the expectations for
future experiments at the NLC with centre--of--mass energies of 500
GeV and 1 TeV, NLC-500  and NLC-1000   and luminosities
of 75 fb$^{-1}$ and 200 fb$^{-1}$, respectively \cite{ABQ2}.  These
figures show  that combining data from the four LEP experiments would
lead to a corresponding bound of $\sim$ 1.9 TeV, while  NLC-500  and
NLC-1000 will allow us to probe sizes  of the order of 8 TeV and 13
TeV.

%%%%%%%%%%%%%%%%%%%%%%%%%%%%%%%%%%%%%%%%%%%%%%%%%%%%%%%%%%%%%%%%%%%
%%%%%%%%%%%%%%%%%%%%%%%%%%%%%%%%%%%%%%%%%%%%%%%%%%%%%%%%%%%%%%%%%%%

%%%%%%%%%%%%%%%%%%%%%%%%%%%%%%%%%%%%%%%%%%%%%%%%%%%%%%%%%%%%%%%%%%%
%%%%%%%%%%%%%%%%%%%%%%%%%%%%%%%%%%%%%%%%%%%%%%%%%%%%%%%%%%%%%%%%%%%

\subsection {Production at hadron colliders}

At collider experiments, there are three different channels 
$l^+ l^-$, $l^{\pm} \nu$ and dijets where exchange of KK excitations of 
photon+$Z$, $W^{\pm}$ and gluons can produce observable deviations from the standard model expectations.

Let's illustrate in details the first case with exchange of neutral bosons.
KK excitations are produced in  
Drell--Yan processes $pp \rightarrow
l^+l^-X$ at the LHC, or $p{\bar p} \rightarrow l^+l^-X$ at the Tevatron, with
$l=e,\mu,\tau$ wich  originate from 
the subprocess $q{\bar q}\rightarrow l^+ l^- X$ of centre--of--mass energy $M$.

The two colliding partons take a fraction
\begin{equation}
x_a={M \over \sqrt s}\ e^{y} \quad{\rm and}\quad
x_b={M \over \sqrt s}\ e^{-y}
\end{equation}
of the momentum of the initial proton ($a$) and (anti)proton ($b$), with a
probability described by the quark or antiquark distribution functions
$f^{(a)}_{q,\bar q}(x_{a}, M^2)$ and $f^{(b)}_{q,\bar q}(x_{b}, M^2)$. The total cross section, due to the production is given by:
\begin{equation}
\sigma= \sum_{q={\rm quarks}} \int^{\sqrt s }_0 dM \int^{\ln (\sqrt s
/M)}_{\ln (M/\sqrt s)}dy \ g_q (y, M) S_q (y, M) \ ,
\end{equation}
where
\begin{equation}
g_q (y, M)= {M \over 18\pi} x_a x_b \ [f^{(a)}_q (x_a,M^2)
f^{(b)}_{\bar q} (x_b, M^2) + f^{(a)}_{\bar q} (x_a, M^2) f^{(b)}_q (x_b,
M^2)]\ ,
\end{equation}
and
\begin{equation}
S_q (y, M)= 
\sum_{\alpha ,\beta\gamma, Z, KK}g^2_{\alpha}(M) g^2_{\beta}(M) 
{(v^{\alpha}_e v^{\beta}_e+
a^{\alpha}_e a^{\beta}_e)(v^{\alpha}_l v^{\beta}_l + a^{\alpha}_l
a^{\beta}_l) \over (s -m^2_{\alpha} + i\Gamma{_\alpha}
m_{\alpha})(s-m^2_{\beta} - i\Gamma_{\beta} m_{\beta})} \ .
\end{equation}

\begin{figure}[htb]
\centering
\epsfxsize=3.5in
\hspace*{0in}
\epsffile{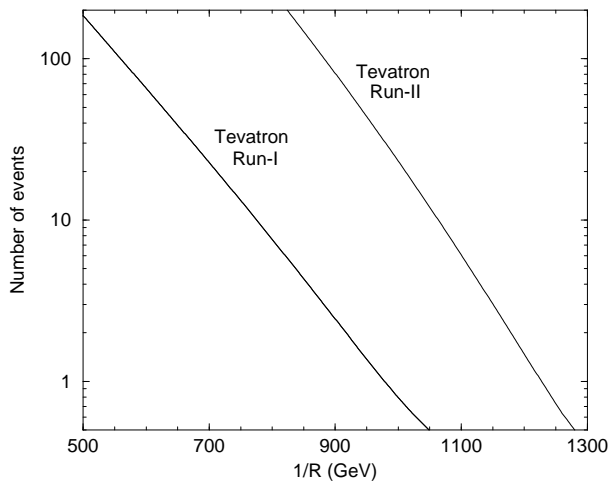}
\caption{\it Number of $l^+ l^-$-pair events with centre--of--mass 
energy above 400 GeV (600 GeV) expected at the Tevatron run-I (run-II) 
with integrated luminosity $\int {\cal L}dt= 110\ pb^{-1}$
($\int {\cal L}dt= 2\ fb^{-1}$) and efficiency times acceptance of
$\sim$ 50\%, as a function of $R^{-1}$.}
\label{fig:fig3}
\end{figure}

 At the Tevatron, the CDF collaboration 
has collected an integrated luminosity $\int {\cal L}dt= 110\ \rm{pb}^{-1}$ 
during the 1992-95 running period. A lower bound on the 
size of compactification scale can be extracted from the absence of candidate 
events at $e^+e^-$ invariant mass above 400 GeV. A similar analysis can be carried over for the case of run-II of the Tevatron with a centre--of--mass energy $\sqrt{s}=2$ TeV and integrated luminosity $\int {\cal L}dt= 2\ fb^{-1}$. The expected number of events at these experiments are plotted in  Fig.~\ref{fig:fig3} while the bounds are summarized in table 1 (the factor $A$ in (\ref{nt}) has
 be  taken to be  50 \%) \cite{ABQ2,AAB}.

 %%%%%%%%%%%%%%%%%%%%%%%%%%%%%%%%%%%%%%%%%%%%%%%%%%%%%%%%%%%%%%%%%%%
%%%%%%%%%%%%%%%%%%%%%%%%%%%%%%%%%%%%%%%%%%%%%%%%%%%%%%%%%%%%%%%%%%%

\begin{figure}[htb]
\centering
\epsfxsize=3.5in
\hspace*{0in}
\epsffile{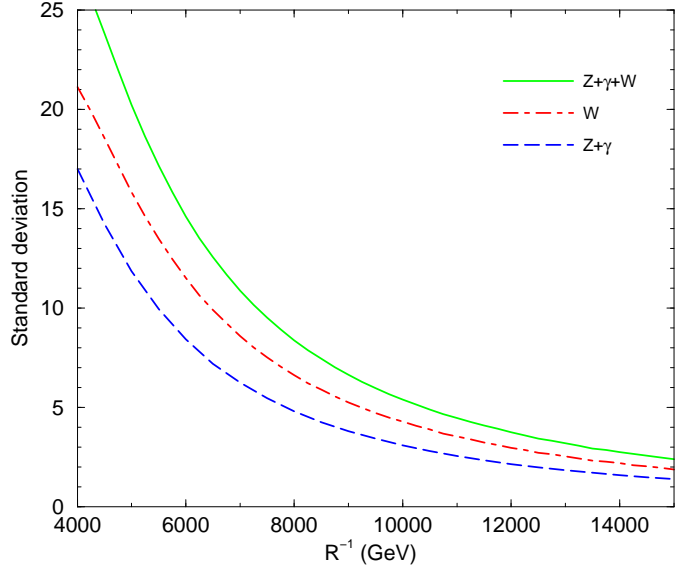}
\caption{\it Number of standard deviation in the number of  $l^+ l^-$
pairs  
and $\nu_l l$ pairs produced from the expected standard model value due to
the presence of one extra-dimension of radius $R$.  }
\label{fig:fig4}
\end{figure}

The most  promising for probing TeV-scale extra-dimensions
are the LHC future experiments at $\sqrt s =14$ TeV with an 
integrated luminosity $\int {\cal L}dt= 100\ fb^{-1}$.  Fig.~\ref{fig:fig4}
shows the expected  deviation from the standard model predictions 
 of the total number of events in the $l^+ l^-$, $l^{\pm} \nu$ due to 
KK excitations $\gamma^{(n)} + Z^{(n)}$ and $W_{\pm}^{(n)}$ respectively.
 The results were obtained by requiring for the dilepton final state 
one lepton to be in  
the central region, $|\eta_l|\le$ 1, the other one having a looser
cut $|\eta_{l^\prime}|\le$ 2.4. Moreover the  lower bound on the transverse 
and invariant mass was chosen to be 400 GeV \cite{AAB,ABQ,ABQ2}.

%%%%%%%%%%%%%%%%%%%%%%%%%%%%%%%%%%%%%%%%%%%%%%%%%%%%%%%%%%%%%%%%%%%
%%%%%%%%%%%%%%%%%%%%%%%%%%%%%%%%%%%%%%%%%%%%%%%%%%%%%%%%%%%%%%%%%%%

\begin{figure}[htb]
\centering
\epsfxsize=3.5in
\hspace*{0in}
\epsffile{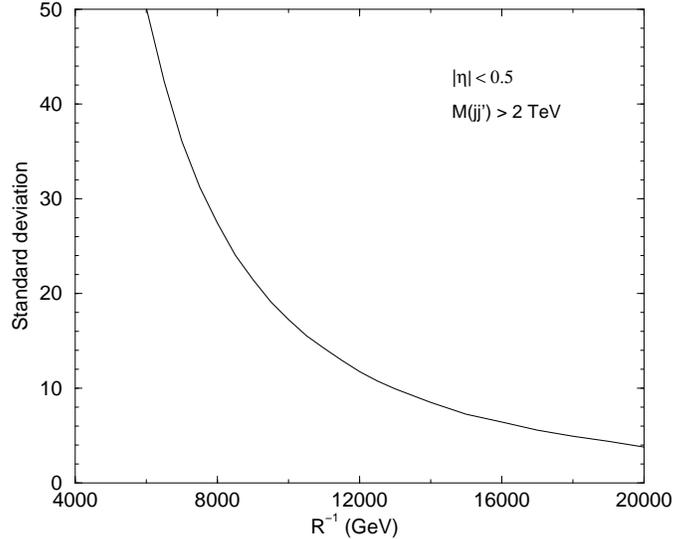}
\caption{\it Number of standard deviation in number of observerd dijets
from 
the expected standard model value, due to the presence of a TeV-scale 
extra-dimension of compactification radius $R$.}
\label{fig:fig5}
\end{figure}

In the case of $(l,l,l)$ scenario, looking for an excess of dijet events
due to KK excitations of gluons could be the most efficient 
channel to constrain the size of extra-dimensions. Fig.~\ref{fig:fig5}
shows the corresponding expected   deviation $\Delta_T$ as defined 
in (\ref{deltaT}). This analysis uses 
summation over all jets,
top excluded,  a rapidity cut, $|\eta |\le$
0.5,
on both jets and requirement on the invariant mass to be $M_{jj'}\ge$ 2
TeV, which reduces the SM background and gives the optimal ratio
$S/\sqrt{B}$ expecially for large masses \cite{AAB}.

%%%%%%%%%%%%%%%%%%%%%%%%%%%%%%%%%%%%%%%%%%%%%%%%%%%%%%%%%%%%%%%%%%%
%%%%%%%%%%%%%%%%%%%%%%%%%%%%%%%%%%%%%%%%%%%%%%%%%%%%%%%%%%%%%%%%%%%

\begin{figure}[htb]
\centering
\epsfxsize=3.5in
\hspace*{0in}
\epsffile{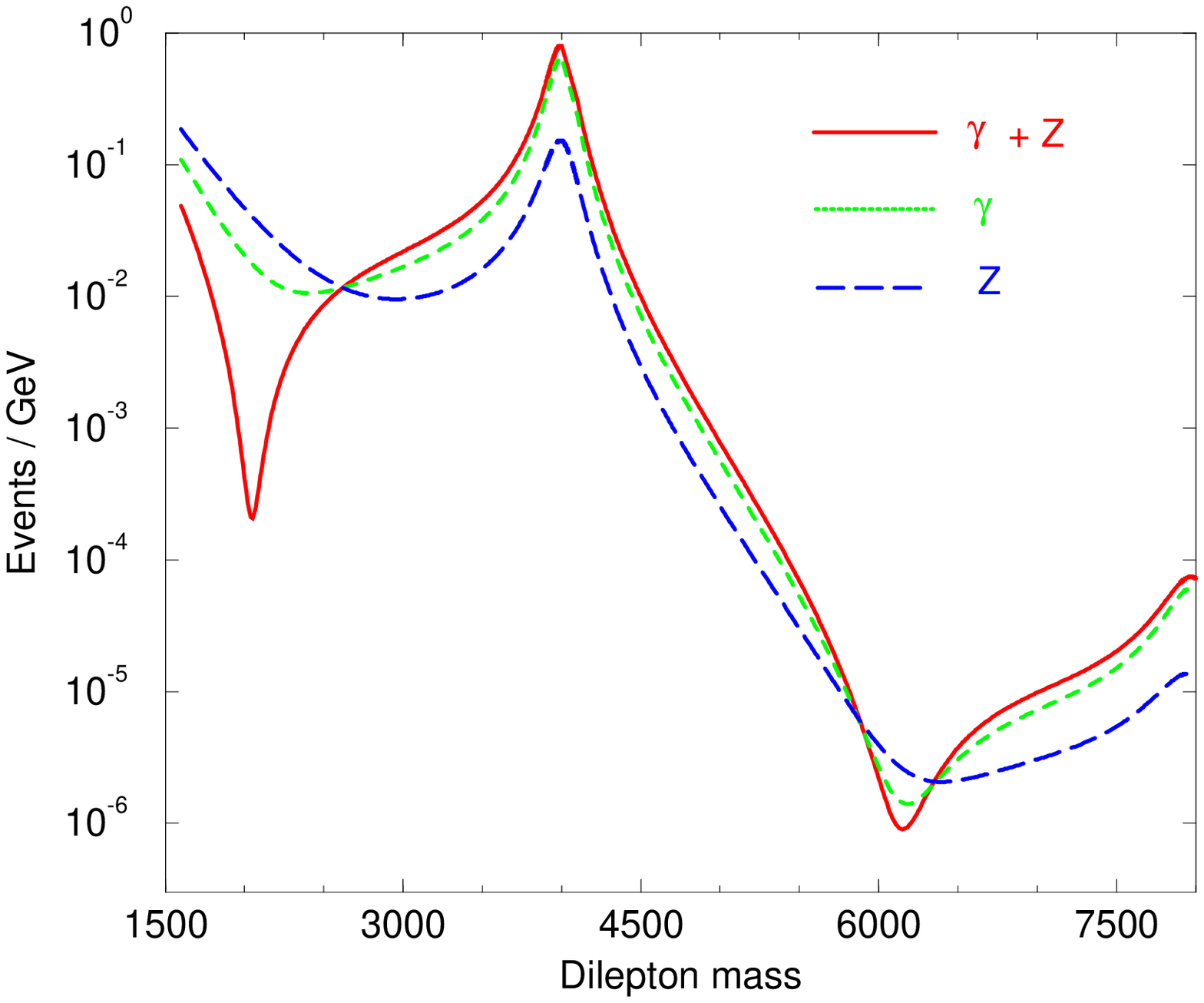}
\caption{\it First resonances in the LHC experiment due to a KK excitation 
of photon and Z  for one extra-dimension at 4 TeV. From highest to lowest: 
excitation of photon+Z, photon and Z boson.}
\label{fig:fig6}
\end{figure}

In addition to these virtual effects, the LHC experiments allow the
production on-shell of KK excitations. The discovery limits for these
KK excitations are given in Table 1. An interesting observation is the
case of excitations $\gamma^{(1)} + Z^{(1)}$ where interferences lead
to a ``deep'' just before the resonance as illustrated in
Fig.~\ref{fig:fig6}
%%%%%%%%%%%%%%%%%%%%%%%%%%%%%%%%%%%%%%%%%%%%%%%%%%%%%%%%%%%%%%%%%%%
%%%%%%%%%%%%%%%%%%%%%%%%%%%%%%%%%%%%%%%%%%%%%%%%%%%%%%%%%%%%%%%%%%%

There are some ways to distinguish the corresponding signals from other
possible origin of new physics, such as models with new gauge bosons. In the
case of observation of resonances, one expects three resonances in  the
$(l,l,l)$ case and two in the  $(t,l,l)$ and $(t,l,t)$ cases, located
practically at the same mass value. This property is not shared by most of
other new gauge boson models. Moreover, the heights and widths of the
resonances are directly related to those of standard model gauge bosons in
the corresponding channels.  In the case of virtual effects, these are not
reproduced by a tail of Bright-Wigner shape and a deep is expected just
before the resonance of the photon+$Z$, due to the interference between the
two. However, good statistics will be necessary\cite{ABQ}.

\subsection {High precision data low-energy bounds}

Using the lagrangian describing interactions of the standard 
model states, it is possible to compute all physical observables
in term of few input data. Then one can compare the predictions with 
experimental values.
 
Following ~\cite{scc,Delgado} we will use as  input parameters,
the Fermi constant $G_F=1.166\times
10^{-5}$ GeV$^{-2}$, the fine-structure constant $\alpha=1/137.036$ 
(or $\alpha(M_Z)=1/128.933$) and the
mass of the $Z$ gauge-boson $M_Z=91.1871$ GeV. The observables 
given in Table 3 are then computed with the new lagrangian including 
the contribution of KK excitations. The effects of the latter will be computed 
as a leading order expansion in the small parameter
\begin{equation}
\label{equis}
X=\sum_{n=1}^\infty \frac{2}{n^2}\, \frac{m_Z^2}{M_c^2}=\frac{\pi^2}{3}
{(m_Z R_\parallel)}^2\, ,
\end{equation}
as one expects $ m_Z \ll 1/R_\parallel $.
 
 Here we 
follow ~\cite{Delgado} and summarize the bounds in the case where all fermions are localized. Other possibilities have been discussed in ~\cite{Delgado}. 
The results depend on the choice of 
Higgs fields  to be  localized or  bulk states.
In  the case where the Higgs  scalar is  localized,
it does not conserve internal momentum and  its vacuum expectation
value could mix different KK excitations with massless gauge bosons.
To include both possibilities of identification of Higgs scalar with
 localized or bulk states, we consider the Standard model with two Higgs 
doublets, $H_i\ (i=1,2)$, with $v^2\equiv\langle H_1\rangle^2+ 
\langle H_2\rangle^2$ and
$v\simeq 174$ GeV, and introduce the following mixing angles:
\bea
\tan\beta &=& \langle H_2\rangle/\langle H_1\rangle \\
\sin^2\alpha &=&\varepsilon^{H_2}\,\sin^2\beta+\varepsilon^{H_1}\,\cos^2\beta
\label{mixing}
\eea
with the operator $\varepsilon$ defined as
$\varepsilon^{\phi}=1\ (0)$ for the field $\phi$ localized or not.

The relevant part of the lagrangian is then:

\begin{equation}
\label{charged}
{\mathcal{L}}= {\mathcal{L}}^{neutral} +\sum_{a=1}^2 {\mathcal{L}}_a^{ch}
\end{equation}
 where the neutral current sector part takes the form:

\bea
\label{neutrals}
{\mathcal{L}}^{neutral}&=&\frac{1}{2}m_Z^2 Z\cdot Z 
+  \frac{1}{2} \sum_{n=1}^{\infty} M^2_{\vec n}\,\left[Z^{(n)}\cdot Z^{(n)}+
A^{(n)}\cdot A^{(n)}\right]\nonumber\\
&+&\sqrt{2}\sin^2{\alpha} m_Z^2  \sum_{n=1}^\infty Z\cdot Z^{(n)} \nonumber\\
&-&\frac{e}{\sin{\theta} \cos{\theta}}\left[Z_\mu +\sqrt{2}\sum_{n=1}^\infty
{Z^{(n)}}_\mu \right]\left[ \sum_\psi \bar{\psi}\,\gamma^\mu 
\left(g_{V}^{\psi}+\gamma_5\,
g_{A}^{\psi} \right)\psi\, \right]\nonumber\\
&-& e\left[A_\mu +\sqrt{2}\sum_{n=1}^\infty A^{(n)}_\mu
\right]\left[ \sum_\psi Q_\psi \bar{\psi}\,\gamma^\mu 
\psi\, \right]\, ,
\eea

while for the charged currents sector:

\begin{eqnarray}
\label{chargeda}
{\mathcal{L}}_{a}^{ch}&=&
\frac{1}{2}m^{2}_W
W_a\cdot W_a
+\frac{1}{2}\sum_{n=1}^\infty \, M^2_{\vec n} \, W_a^{(n)}\cdot W_a^{(n)}
\nonumber\\ 
&+& m^{2}_W\sqrt{2} \sin^2 {\alpha} \sum_{n=1}^\infty
W_a\cdot W_a^{(n)}
\nonumber\\
&-&g\, \left[ {W_a}_\mu -\,\sqrt{2}  \sum_{n=1}^\infty {W_a^{(n)}}_\mu\right]  \left[ \sum_\psi \bar{\psi}_L \gamma^\mu \frac{\sigma_a}{2}\psi_L \right]
\end{eqnarray}

with $m^{2}_W=g^2v^2/2$,  $m^{2}_Z=(g^2+g^{\prime\, 2})v^2/2$,
$\theta$ is the weak mixing angle defined by the usual relation $e=g\,
\sin \theta=g'\, \cos \theta$ . For energies much below the
electroweak scale the gauge bosons can be integrated out leading to
effective four-fermion operators:

\bea
\label{lowneutral}
{\mathcal{L}}_{low}&=&-\,\frac{1}{2 \left[1- \sin^4{\alpha}\, X \right]m_Z^2}\,
\frac{e^2}{s^2_\theta\, c^2_\theta}\,
\left[
\left(1- \sin^2\alpha\, X \right)^2  - X \right]{\left[ \sum_\psi \bar{\psi}\,\gamma^\mu 
\left(g_{V}^{\psi}+\gamma_5\,
g_{A}^{\psi} \right)\psi\, \right]^2} \nonumber\\
&-&e A^\mu \left[ \sum_\psi Q_\psi \bar{\psi}\,\gamma^\mu 
\psi\, \right]\nonumber\\
&-&X \frac{e^2}{2\, \left[1- \sin^4\alpha \,
X \right]m_Z^2}\, \left[ \sum_\psi Q_\psi \bar{\psi}\,\gamma_\mu 
\psi\, \right]\left[ \sum_\psi Q_\psi \bar{\psi}\,\gamma^\mu 
\psi\, \right] \\
&-&\frac{g^2}{2 M_W^2}\left\{
\left[1-\sin^2\alpha\, \cos^2\theta\,
X \right]^2 
+\, \cos^2\theta \,X \right\} \left[\sum_\psi \bar{\psi}_L \gamma_\mu \frac{\sigma_a}{2}\psi_L\right]  \left[\sum_\psi \bar{\psi}_L \gamma^\mu \frac{\sigma_a}{2}\psi_L\right] \nonumber
\eea

Using this lagrangian it is possible to extract prediction for the 
LEP (high-energy) observables:

\bea {\displaystyle
M_W^2}&=&\left(M_W^2\right)^{SM}\left[1-\frac{\sin^2{\theta}}{\cos{2\theta}}
(\cos{2\alpha} \cos^2{\theta} - \sin^4{\alpha}\sin^2{\theta} +
\sin^4{\alpha}\cos{2\theta}) \, X\right] \nonumber\\&& \nonumber\\
{\displaystyle \frac{\Gamma_{\ell\,\ell}}{\Gamma_{\ell\,\ell}^{SM}}
}&=& \left[1- (\cos{2\alpha} \cos^2{\theta} -
\sin^4{\alpha}\sin^2{\theta} -2 \sin^2{\alpha} ) \,  X \right]\,
, \nonumber\\ && \nonumber\\ {\displaystyle
\frac{\Gamma_{had}}{\Gamma_{had}^{SM}} }&=&\left[1- (\cos{2\alpha}
\cos^2{\theta} - \sin^4{\alpha}\sin^2{\theta} -2 \sin^2{\alpha} ) \,
X\right]\, ,\\&& \nonumber\\  {\displaystyle
\frac{A_{FB}^\ell}{A_{FB}^{\ell,\,SM}} }&=&1 \, ,\nonumber
\end{eqnarray}
and for the low energy observables:
\begin{equation}
\label{QW}
Q_W=\left[ 1-(\cos{2\alpha} \cos^2{\theta} - \sin^4{\alpha}\sin^2{\theta}) X 
\right]\, Q_W^{SM}+16\, \delta\, Q_W\, ,
\end{equation}
where
\begin{eqnarray}
\label{deltaQW}
\delta Q_W&=&-\frac{1}{4}\, \frac{s^2_{\theta} \, c^2_{\theta}}
{c_{2\theta}}\, Z\, (\cos{2\alpha} \cos^2{\theta} - \sin^4{\alpha}\sin^2{\theta}) X\nonumber\\
&+&\,X\, \left\{ (2Z+N)
\left[g_A^{e} g_V^{u}-\frac{1}{4}
\sin^2\alpha \,
g_V^{u}-
\sin^2\alpha \,
\left(\frac{1}{4}-\frac{2}{3}
\sin^2{\theta}\right)g_A^{e}\right]\right.\nonumber\\
&+&\left.(Z+2N)\left[ g_A^{e} g_V^{d}-\frac{1}{4}
\sin^2\alpha \, 
g_V^{d}
-\sin^2\alpha \,
\left(-\frac{1}{4}+\frac{1}{3}
\sin^2{\theta}\right) g_A^{e} \right]
\right\}
\end{eqnarray}

The experiments are carried with Cesium atoms for which the number of protons 
and neutrons are $Z=55$ and
$N=78$,  while $\sum_{q'}\left|V_{uq'}\right|^2 = 1$ does not receive
corrections at the leading order.  Performing a $\chi^2$ fit, one finds
for example that   if the Higgs is assumed to be a bulk state ($\alpha =0$) 
like the
gauge bosons, $R^{-1} \simgt 3.5$ TeV. Inclusion of $Q_W$ measurement,
which does not give a good agreement with the  standard model itself,
raises the bound to $R^{-1} \simgt 3.9$ TeV~\cite{Delgado}. Different choices 
for localization of matter states and  Higgs  lead to slightly different bounds, lying in the 1 to 5 TeV range, and the analysis can  be found  
in ~\cite{Delgado}.

\subsection {One extra dimension for other cases:}

Except for the  $(l,l,l)$ scenario, in all other cases there are no
excitations of gluons and there no important limits from the dijets
channels \cite{AAB}.

The KK excitations  $W_{\pm}^{(n)}$, $\gamma^{(n)}$ and $Z^{(n)}$ are
present 
and lead to the same limits in the $(t,l,l)$ case:  6 TeV for discovery and
15 TeV for the exclusion bounds.

In the $(t,l,t)$ case, only the $SU(2)$ factor feels the
extra-dimension  and the limits are set by KK excitations of $W^{\pm}$
and are again 6 TeV for discovery and 14 TeV for the exclusion bounds.

\begin{figure}[htb]
\centering
\epsfxsize=3.5in
\hspace*{0in}
\epsffile{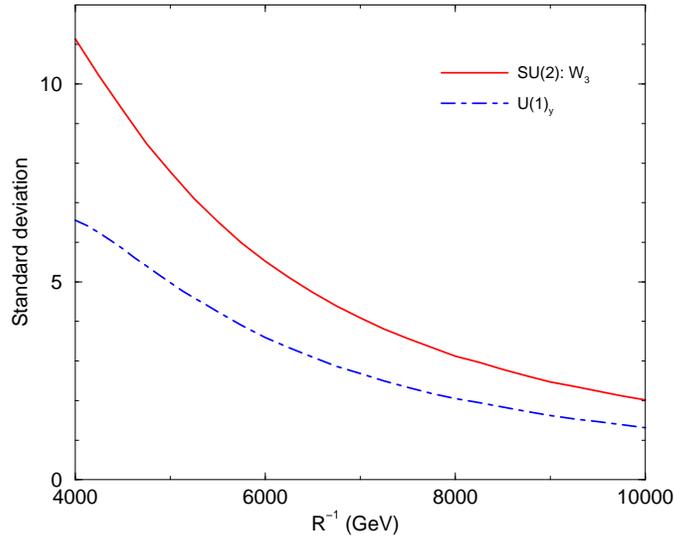}
\caption{\it Number of standard deviation in the number of  $l^+ l^-$
pairs  
produced from the expected standard model value due to the presence of 
one extra-dimension of radius $R$ in the case of $(t,l,t)$ i.e. $W_3$ 
KK excitations and the case
of $(t,t,l)$ i.e. $U(1)_Y$ boson KK excitations.  }
\label{fig:fig7}
\end{figure}

In the $(t,t,l)$ channel where only $U(1)_Y$ feels 
the extra-dimension the limits are
weaker, the exclusion bound is in fact around 8 TeV, as can be seen in
Fig.~\ref{fig:fig7}.

%%%%%%%%%%%%%%%%%%%%%%%%%%%%%%%%%%%%%%%%%%%%%%%%%%%%%%%%%%%%%%%%%%%
%%%%%%%%%%%%%%%%%%%%%%%%%%%%%%%%%%%%%%%%%%%%%%%%%%%%%%%%%%%%%%%%%%%

In addition to these simple possibilities, brane constructions lead often 
to cases where part of $U(1)_Y$
is $t$ and part is $l$, while $SU(3)$ and $SU(2)$ are either $t$ or $l$.
If $SU(3)$ is  $l$ then the bounds come from dijets, if instead $SU(3)$ is  
$t$ and $SU(2)$ is $l$ the limits could come from $W^{\pm}$ while if both are 
$t$ then it will be difficult to distinguish this case 
from a generic extra $U(1)'$.  A good
statistics would be needed to see the deviation in the tail of the
resonance as being due to effects additional to those of a generic  $U(1)'$
resonance.

\subsection {More than one extra dimensions}

The computation of virtual effects of KK excitations involves summing on effects of a priori infinite number of tree-level diagrams as terms of the form:
\begin{equation}
\sum_{|\vec{n}|} \frac {g^2{(|\vec{n}|)}}{|\vec{n}|^2}
\label{sumdiv}
\end{equation}
arising  from interference  between the exchange of the photon and
$Z$-boson and their KK excitations, with $g^2(|\vec{n}|)$ the KK-mode 
couplings. In the case of one extra-dimension the sum in (\ref{sumdiv})  
converges rapidly and for $R M_s \sim {\cal O}(10)$ the result is not sensitive to the value of $M_s$. This alowed us to discuss bounds on only one paprameter, the scale of compactification. 

In the case of two or more dimensions, Eq. (\ref{sumdiv}) is 
divergent and needs to be regularized using: 
\begin{equation}
g(|\vec{n}|)\sim g\; a_{(|\vec{n}|)}\; 
e^{\frac {-c|\vec{n}|^2}{2R^2 M_s^2}}\, ,
\end{equation}
where $c$ is a constant and  $a_{(|\vec{n}|)}$  takes into account the
normalization of the gauge kinetic terms, as only  the even combination
couples to the boundary. For the case of two  extra-dimensions \cite{ABQ2}
$a_{(0,0)} =1$, $a_{(0,p)} =a_{(q,0)}=\sqrt{2}$ and  $a_{(q,p)} =2$
with $(p,q)$ positive ($>0$) integers. The result  will depend on both
the compactification and string scales.  Other features are that
cross-sections are bigger and  resonances are closer.  The former property
arises because the degeneracy of states  within each mass level
increases with the number of  extra dimensions while the latter
property implies  that more resonances  could be reached by a given hadronic
machine.

\section{Extra-dimensions transverse to the world brane: KK excitations
of gravitons}\label{subsec:miss}

The localization of (infinitely massive) branes in the $(D-d)$ dimensions 
breaks translation invariance  along these directions. Thus, the
corresponding momenta are not conserved: particles, as gravitons, could be
absorbed or emitted from the brane into the $(D-d)$ dimensions. Non
observation of the effects of such processes  allow us to get bounds on the
size of these transverse extra dimensions. In order  to simplify the
analysis, it is usually assumed that among the $D-d$ dimensions $n$ have
very large common radius $R_{\perp } \gg M_s^{-1}$, while the remaining
$D-d-n$ have sizes of the order of the string length.

\subsection{Signals from missing energy experiments}

During a collision of center of mass energy $\sqrt{s}$, there are 
$(\sqrt{s}R_{\perp})^n$ KK excitations of gravitons with mass
$m_{KK\perp}<\sqrt{s}< M_s$, which can be emitted. Each of these states 
looks from the four-dimensional point of view as a massive, quasi-stable, 
extremely weakly coupled ($s/M^2_{pl}$ suppressed) particle that escapes
from the detector. The total effect is a missing-energy cross section
roughly of order: 
\be
\frac {(\sqrt{s}R_{\perp })^n} {M^2_{pl}} \sim \frac{1}{s} 
{(\frac{\sqrt{s}}{M_s})^{n+2}}
\label{miss1}
\ee

For illustration, the simplest process is the gluon annihilation into a
graviton which escapes into the extra dimensions. The corresponding
cross-section is given by (in the weak coupling limit)~\cite{aadd}:
\be
\sigma(E)\sim {E^n\over M_s^{n+2}}{\Gamma\left(1-2E^2/M_s^2\right)^2
\over\Gamma\left(1-E^2/M_s^2\right)^4}\, ,
\label{sigma}
\ee
where $E$ is the center of mass energy and $n$ the number of extra large
transverse dimensions. The above expression exhibits 3 kinematic regimes
with different behavior. At high energies $E\gg M_s$, it falls off
exponentially due to the UV softness of strings. At energies of the order
of the string scale, it exhibits a sequence of poles at the position of
Regge resonances. Finally, at low energies $E\ll M_s$, it falls off as a
power $\sigma(E)\sim E^n/M_s^{n+2}$, dictated by the effective higher
dimensional gravity which requires the presence of the
$(4+n)$-dimensional Newton's constant $G_N^{(4+n)}\sim l_s^{n+2}$
from eq.(\ref{GN}).

Explicit computation of these effects leads to the bounds given in table 2~\cite{missing}.
The results require some remarks:
\begin{itemize}

\item The amplitude for emission of each of the KK gravitons is taken to be
well approximated by the tree-level coupling of the massless graviton as
derived from General Relativity. Eq.~\ref{coupling} suggests that this is
likely to be a good approximation for $R_{\perp}M_s\gg 1$.

\item The cross-section depends on the size $R_\perp$ of the transverse
dimensions and allows to derive bounds on this {\it physical} scale. 
As it can be seen from Eq.~(\ref{coupling}), transforming these bounds to 
limits on $M_s$ there is an ambiguity on different factors involved, such as
the string  coupling. This is sometimes absorbed in the so called
``fundamental quantum gravity scale $M_{(4+n)}$''.  Generically $M_{(4+n)}$
is bigger than $M_s$, and in some cases, as in type II strings or in
heterotic strings with small instantons, it can be many orders of magnitude
higher than $M_s$. It  corresponds to the scale  where the 
perturbative expansion of string theory seems to break down \cite{veneziano}.

\item There is a particular energy and angular distribution  of the
produced gravitons that arises from the distribution in mass  of KK
states given in  Eq.~(\ref{KKdef}). It might be a smoking gun for the
extra-dimensional nature of such observable signal.

\item For given value of $M_s$, the cross section for graviton emission
decreases with the number of large transverse dimensions. The effects
are more likely to be observed for the lowest values of $M_s$ and $n$.

\item Finally, while the obtained bounds for $R_\perp^{-1}$ are 
smaller than those that could be checked in table-top experiments probing 
macroscopic gravity at small distances (see next subsection), 
one should keep in mind that 
larger radii are allowed if one relaxes the assumption of isotropy, 
by taking for instance two large dimensions with different radii.

\end{itemize}

In table 2, we have also included astrophysical and cosmological
bounds. Astrophysical bounds~\cite{astcos,supernovae} arise from the
requirement that the radiation of gravitons should not carry on too much
of the gravitational binding energy released during core collapse of
supernovae. In fact, the measurements of Kamiokande and IMB for SN1987A 
suggest that the main channel is neutrino fluxes.

The best cosmological bound~\cite{COMPTEL} is obtained from requiring that
decay of bulk gravitons to photons do not generate a spike in the energy
spectrum of the photon background measured by the COMPTEL instrument. Bulk 
gravitons  are  expected  to be produced just before
nucleosynthesis due to thermal radiation from the brane. The limits assume
that the temperature was at most 1 MeV as nucleosynthesis begins, and become
stronger if the temperature is increased.

\subsection{Gravity modification and sub-millimeter forces}

Besides the spectacular experimental predictions in particle
accelerators, string theories with large volume compactifications and/or
low string scale predict also possible modifications of gravitation in
the sub-millimeter range, which can be tested in ``tabletop" experiments
that measure gravity at short distances. There are two categories of such
predictions:\hfil\\  
(i) Deviations from the Newton's law $1/r^2$ behavior to $1/r^{2+n}$, for
$n$ extra large transverse dimensions, which can be observable for $n=2$ 
dimensions of sub-millimeter size. This case is particularly attractive
on theoretical grounds because of the logarithmic sensitivity of Standard
Model couplings on the size of transverse space, but also for
phenomenological reasons since the effects in particle colliders are
maximally enhanced~\cite{missing}. Notice also the coincidence of this
scale with the possible value of the cosmological constant in the
universe that recent observations seem to support.\hfil\\
(ii) New scalar forces in the sub-millimeter range, motivated by the
problem of supersymmetry breaking discussed in section {\it 4.3}, and
mediated by light scalar fields $\varphi$ with
masses~\cite{fkz,iadd,aadd,ads}:
\be
m_{\varphi}\simeq{m_{susy}^2\over M_P}\simeq 
10^{-4}-10^{-2}\ {\rm eV} \, ,
\label{msusy}
\ee
for a supersymmetry breaking scale $m_{susy}\simeq 1-10$ TeV. These
correspond to Compton wavelengths in the range of 1 mm to 10 $\mu$m.
$m_{susy}$ can be either the KK scale $1/R$ if supersymmetry is broken by
compactification~\cite{iadd}, or the string scale if it is broken
``maximally" on our world-brane~\cite{aadd,ads}. A model independent
scalar force is mediated by the radius modulus (in Planck units)
\be
\varphi\equiv\ln R\, ,
\label{varphi}
\ee
with $R$ the radius of the longitudinal or transverse dimension(s),
respectively. In the former case, the result (\ref{msusy}) follows from
the behavior of the vacuum energy density $\Lambda \sim 1/R^4$ for large
$R$ (up to logarithmic corrections). In the latter case, supersymmetry is
broken primarily on the brane only, and thus its transmission to the bulk
is gravitationally suppressed, leading to masses (\ref{msusy}).

The coupling of these light scalars to nuclei can be computed since it
arises dominantly through the radius dependence of $\Lambda_{\rm QCD}$,
or equivalently of the QCD gauge coupling. More precisely, the coupling
$\alpha_\phi$ of the radius modulus (\ref{varphi}) relative to gravity
is~\cite{iadd}:
\be
\alpha_\varphi = {1\over m_N}{\partial m_N\over\partial\varphi}=
{\partial\ln\Lambda_{\rm QCD}\over\partial\ln R}= 
-{2\pi\over b_{\rm QCD}}{\partial\over\partial\ln R}\alpha_{\rm QCD}\, ,
\label{dcoupling}
\ee
with $m_N$ the nucleon mass and $b_{\rm QCD}$ the one-loop QCD
beta-function coefficient. In the case where supersymmetry is broken 
primordially on our
world-brane at the string scale while it is almost unbroken the bulk, the
force (\ref{coupling}) is again comparable to gravity in theories with
logarithmic sensitivity on the size of transverse space, i.e. when there
is effective propagation of gravity in $d_\perp=2$ transverse dimensions.
The resulting forces can therefore be within reach of upcoming
experiments~\cite{price}.

In principle there can be other light moduli which couple with even
larger strengths. For example the dilaton $\phi$, whose VEV determines
the (logarithm of the) string coupling constant, if it does not acquire
large mass from some dynamical supersymmetric mechanism, can lead to a
force of strength 2000 times bigger than gravity~\cite{tvdil}.

%%%%%%%%%%%%%%%%%%%%%%%%%%%%%%%%%%%%%%%%%%%%%%%%%%%%%%%%%%%%%%%%%%%
\begin{figure}[htb]
\centering
\epsfxsize=3in
\hspace*{0in}
%\vspace*{1in}
\epsffile{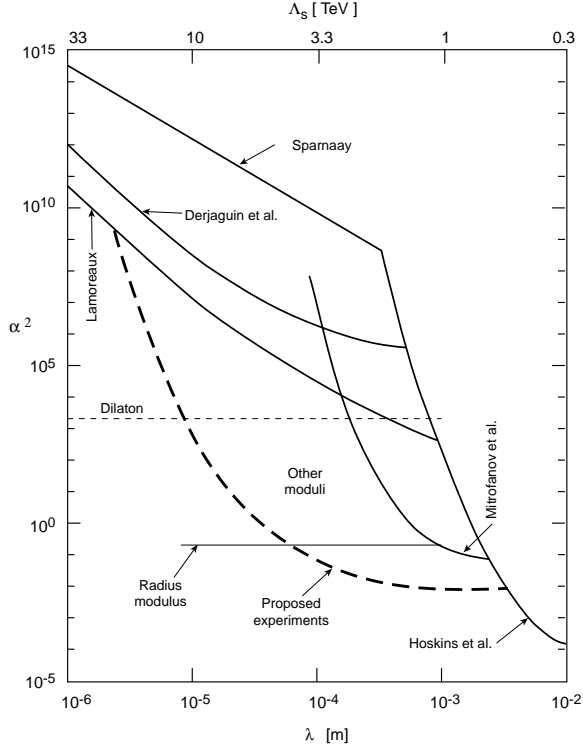}
\caption{\it Strength of the modulus force relative to gravity ($\alpha^2$)
versus  its Compton wavelength ($\lambda$).}
\label{fig:fig8}
\end{figure}
%%%%%%%%%%%%%%%%%%%%%%%%%%%%%%%%%%%%%%%%%%%%%%%%%%%%%%%%%%%%%%%%%%%

In fig.~\ref{fig:fig8} we depict the actual information from previous,
present and upcoming experiments~\cite{price}. The vertical axis is the
strength, $\alpha^2$, of the force relative to gravity; the horizontal
axis is the Compton wavelength of the exchanged particle; the upper scale
shows the corresponding value of the supersymmetry breaking scale (large
radius or string scale) in TeV. The solid lines indicate the present
limits from the experiments indicated. The excluded regions lie above
these solid lines. Measuring gravitational strength forces at such short  
distances is quite challenging. The most important background is the Van
der Walls force which becomes equal to the gravitational force between
two atoms when they are about 100 microns apart. Since the Van der Walls
force falls off as the 7th power of the distance, it rapidly becomes
negligible compared to gravity at distances exceeding 100 $\mu$m. The
dashed thick line gives the expected sensitivity of the present and
upcoming experiments, which will improve the actual limits by roughly two
orders of magnitude and --at the very least-- they will, for the first
time, measure gravity to a precision of 1\% at distances of $\sim$ 100
$\mu$m.

\section{Dimension-eight operators and limits on the string scale}

At low energies, the interaction of light (string) states is
described by an effective field theory. Non-renormalizable dimension-six
operators are due to the exchange of KK excitations of gauge bosons between
localized states. If these are absent, then there are deviations to the
standard model expectations from dimension-eight operators. There are two
generic sources for such operators: exchange of virtual KK excitations of
bulk fields (gravitons,...) and form factors due to the extended nature of
strings.

The exchange of virtual KK excitations of bulk gravitons is described
in the effective  field theory by an amplitude involving the sum
$\frac {1}{M_p^2}\sum_n \frac {1}{s-\frac{{\vec n}^2}{R_\perp^2}}$. For
$n > 1$, this sum diverges and one cannot compute it in field theory but 
only in a fundamental (string) theory. In analogy with  the case of
exchange of gauge bosons, one expects the string scale to act as a cut-off
with the result: 
\be  
A g_s^2 \frac {T_{\mu \nu}T^{\mu \nu} - \frac {1}{1+d_\perp} 
T_\mu^\mu  T_\nu^\nu} {M_s^4}\, .
\label{QFT}
\ee 
The approximation $A = \log{ \frac{M_s^2}{s}}$ for $d_\perp =2$ and 
$A = \frac{2}{d_\perp-2}$ for $d_\perp > 2$ is usually used for 
quantitative discussions. There are some reasons which might invalidate 
this approximation in general. In fact, the result is 
very much model dependent: in type I string models it reflects the 
ultraviolet behavior of open string one-loop diagrams which are model 
(compactification) dependent.

In order to understand better this issue, it is important to remind that in
type I string models, gravitons and other bulk particles correspond to
excitations of closed strings. Their tree-level exchange is described  by a
 cylinder joining the initial $|B in>$ and final $|B out>$ closed
 strings lying on the brane. This cylinder can be be seen on the other
hand as an open string with one of its end-points describing the
 closed (loop)  string $|B in>$, while the other end draws $|B out>$. In
other words, the cylinder can be seen as an annulus which is a one-loop
 diagram of open strings with boundaries $|B in>$ and $|B out>$. Note that
usually the theory requires the presence of other
weakly interacting closed strings besides gravitons.

More important is that when the gauge degrees of freedom
arise from Dirichelet branes, it is expected that the dominant source
of dimension-eight operators is not the exchange of KK states but
instead the effects of massive open string oscillators~\cite{AAB,string}. 
These give rise to
contributions to tree-level scatterings that behave as $g_s s/M_s^4$.
Thus, they are enhanced by a string-loop factor $g_s^{-1}$ compared to
the field theory estimate based on KK graviton exchanges. Although the
precise value of $g_s$ requires a detail analysis of threshold
corrections, a rough estimate can be obtained by taking 
$g_s\simeq\alpha\sim 1/25$, implying an enhancement by one order of
magnitude, and in any case a loop-factor as consequence of perturbation 
theory. 

What is the simplest thing one could do in practice?.  There are some
processes for which there is only one allowed dimension-eight operator; an
example is $f{\bar f}\rightarrow \gamma\gamma$. The coefficient of this
operator can then be computed in terms of $g_s$ and $M_s$.
As a result, in the only framework where computation of such operators is
possible to carry out, one cannot rely on the effects of exchange of KK
graviton excitations in order to derive bounds on extra-dimensions or the
string scale. Instead, one can use the dimension-eight operator arising
from stringy form-factors.

Under the assumption that the electrons arise as open strings on a $D3$-brane, and not as living on the intersections of different kind of branes, an estimate at the lowest order approximation of string form factor in 
type I  was used in \cite{string}. For instance for $e^+ e^-\to \gamma\gamma$
one has:
\be
 {d \sigma\over d \cos\theta} =
         {d \sigma\over d \cos\theta}\biggr|_{SM} \cdot
\left[ 1 + {\pi^2\over 12 } {ut\over M_S^4}  +
                 \cdots \right]  
\label{photphot}
\ee
while for Bhabha scattering,
it was suggested that
\be
   {d \sigma\over d \cos\theta} (e^-e^+\to e^-e^+)  =
         {d \sigma\over d \cos\theta}\biggr|_{SM} \cdot
\left| {\Gamma(1-\frac {s}{M_s^2}) \Gamma(1-\frac {t}{M_s^2}) \over
\Gamma(1-\frac {s}{M_s^2} -\frac {t}{M_s^2})} \right|^2 \ ,
\label{bhabha}
\ee
where $s$ and $t$ are the Mandelstam kinematic variables.
Using these estimates, present LEP data lead to limits on the string scale 
$M_s \simgt 1$ TeV. This translates into a stronger bound on the size of transverse dimension than those obtained from missing energy experiments in the cases $d_\perp > 2$.

\section {Conclusions}

The theoretical possibility of a string scale lying at energies much
lower  than the four-dimensional Planck scale opens up a new insight
on problems of physics beyond the standard model.  Moreover it calls
for two urgent investigations: one is to derive limits on the
possible scales of new physics associated with compactification and
string scales from present experiments, and the second is to understand 
possible signatures of this new  physics at future experiments.

We have stressed that addressing these issues requires a well defined
 theoretical  framework, for instance the choice of a vacuum of string
 theory. One of the issues which illustrates this necessity is for
 instance four-fermion operators which  are due to exchanges of
 virtual KK excitation of the graviton. In addition  to providing new
 bounds on the size of transverse dimensions to the brane, these  effects
 were expected to lead to observable deviations in cross-sections with 
 an angular dependence typical of the exchange of spin-two states. This would
 provide us with a spectacular signature of quantum gravitational
 effects. However such  effects can only be analysed in a consistent theory 
 of quantum gravity: string theory. Unfortunately, carrying such an analysis
shows that in fact these higher dimensional operators are dominated by string 
form factors due to excitation of oscillation modes of the strings  and 
 thus one does not expect to measure effects due to virtual exchange of 
 gravitons. 

 Understanding features of string vacua and building realistic string models 
 will certainly shed some light on some other issues which were not 
 covered by this review. We can cite for instance:

\begin {itemize}

\item Lowering  the string scale, one increases the strength of higher
(non-renormalizable) operators leading to the possibility of inducing
exotic processes at experimentally excluded rates, as for proton decay 
and flavour changing processes. Although an explicit 
string realization of the scenario is necessary in order to have a
satisfactory solution, at the effective field 
theory level many discrete or
global symmetries can be displayed that forbid these operators.

\item As we have stressed in section 3,  string vacua 
 with arbitrarely low string scale is a consequence of the existence of 
 mechanisms ($D$-branes) for localizing gauge degrees of freedom. Is it 
 also possible  to localize gravitons? Starting with 5-dimensional gravity, 
 it was shown in \cite{RS} that there exist metrics, a slice of Anti-de-Sitter  space where gravity is localized. The possible application for 
 phenomenological
 purposes of such a scenario necessites then understanding the origin
 of the other (gauge and matter) states. For instance, if these are localized 
 on a brane at the boundary of this space, then what is the nature of this 
 brane and of the theory leaving on its world-volume? and what kind of 
 modifications (or extensions if any) of the original proposal of \cite{RS}
 are needed in order to have a consistent embedding in string theory? 
 These issues are related to each other 
 and they constitute a intensive subject of research at the present time.

\end{itemize}

\section*{Acknowledgments}
This work was partly supported by the EU under TMR contract ERBFMRX-CT96-0090.
We would like to thank E. Accomando, Y. Oz and M. Quir\'os for enjoyable
collaborations and A. Pomarol for discussions.

%%%%%%%%%%%%%%%%%%%%%%%%%%%%%%%%%%%%%%%%%%%%%%%%%%%%%%%%%%%%%%%%%%%%%
\begin{table}[t]
\caption{Limits on $R^{-1}_\parallel$ in TeV at present and future
colliders. The luminosity  is given in  fb$^{-1}$.\label{tab:1}}
\vspace{0.4cm}
\begin{center}
\begin{tabular}{  | c | c | c | c | l |} 
\hline
  & & & &
\\   Collider & Luminosity &  Gluons  & $W^{\pm}$ & $\gamma + Z$   \\ 
\hline\hline
\mco{5}{|c|}{Discovery of Resonances}   \\ \hline
  LHC     & 100        &  5  & 6  & 6               \\ \hline\hline
\mco{5}{|c|}{Observation of Deviation}   \\ \hline
 LEP 200    &$4\times 0.2$  & - & -  & 1.9 \\ \hline
 TevatronI & $0.11$ &  -  & - & 0.9  \\ \hline 
 TevatronII & 2 &  -  & - & $1.2$ \\ \hline 
  TevatronII  & 20 &  4 & - & 1.3 \\ \hline 
 LHC & 10& 15   & 8.2  & 6.7 \\ \hline
 LHC & 100& 20   & 14 &  12  \\ \hline
  NLC500 & 75& - & - & 8  \\ \hline
  NLC1000 & 200& - & - & 13  \\ \hline
\end{tabular}
\end{center}
\end{table}
%%%%%%%%%%%%%%%%%%%%%%%%%%%%%%%%%%%%%%%%%%%%%%%%%%%%%%%%%%%%%%%%%%%%%%%%

%%%%%%%%%%%%%%%%%%%%%%%%%%%%%%%%%%%%%%%%%%%%%%%%%%%%%%%%%%%%%%%%%%%%%
\begin{table}[t]
\caption{Limits on $R_\perp$ in mm from missing-energy
processes.\label{tab:exp3}}
\vspace{0.4cm}
\begin{center}
\begin{tabular}{  | c | c | c | l |} 
\hline
  & & &
\\   Experiment & $R_\perp (n=2)$ & $R_\perp (n=4)$ & $R_\perp (n=6)$ \\ 
\hline\hline
\mco{4}{|c|}{Collider bounds}   \\ \hline

 LEP 2   & $4.8\times 10^{-1}$ & $1.9\times 10^{-8}$  & 
                              $6.8 \times 10^{-11}$ \\ \hline
  Tevatron  &   $5.5 \times 10^{-1}$  & $1.4 \times 10^{-8}$ 
              & $4.1 \times 10^{-11}$ \\ \hline 
  LHC &  $4.5 \times 10^{-3}$   & $5.6\times 10^{-10}$  & 
                              $2.7 \times 10^{-12}$  \\ \hline
  NLC & $1.2\times 10^{-2}$  & $1.2\times 10^{-9}$  & 
                              $6.5 \times 10^{-12}$  \\ \hline\hline
\mco{4}{|c|}{Present non-collider bounds}   \\ \hline
  
SN1987A   &  $3 \times 10^{-4}$   & 
           $1 \times 10^{-8}$ 
                 & $6 \times 10^{-10} $ \\ \hline
COMPTEL &  $5 \times 10^{-5}$   & - & 
                              - \\ \hline
\end{tabular}
\end{center}
\end{table}
%%%%%%%%%%%%%%%%%%%%%%%%%%%%%%%%%%%%%%%%%%%%%%%%%%%%%%%%%%%%%%%%%%%%%%%%

%
%%%%%%%%%%%%%%%%%%%%%%%%%%%%%%%%%%%%%%%%%%%%%%%%%%%%%%%%%%%%%%%%%%%%%%%%
\begin{table}[t]
\caption{Set of physical observables.  The Standard Model 
predictions are computed for a Higgs mass $M_H=M_Z$ ($M_H=300$ GeV) and a 
top-quark mass $m_t=173\pm 4$ GeV.\label{tab:2}}
\vspace{0.4cm}
\begin{center}
\begin{tabular}{|c|c|c|}\hline & & \\
Observable & Experimental value & Standard Model prediction \\ \hline
\hline $M_W$ (GeV) & 80.394$\pm$0.042 & 80.377$\pm$0.023 ($-$0.036)\\
$\Gamma_{\ell\ell}$ (MeV) & 83.958$\pm$0.089 & 84.00$\pm$0.03 ($-$0.04)\\
$\Gamma_{had}$ (GeV)& 1.7439$\pm$0.0020 & 1.7433$\pm$0.0016 ($-$0.0005)\\
$A_{FB}^\ell$& 0.01701$\pm$0.00095 & 0.0162$\pm$0.0003 ($-0.0004$)\\
$Q_W$ & $-$72.06$\pm$0.46 & $-$73.12$\pm$0.06 ($+$0.01)\\
$\sum_{i=1}^{3}\left|V_{1i}\right|^2$ & 0.9969$\pm$0.0022 & 1 (unitarity)\\
\hline 
\end{tabular}
\end{center}
\end{table}
%%%%%%%%%%%%%%%%%%%%%%%%%%%%%%%%%%%%%%%%%%%%%%%%%%%%%%%%%%%%%%%%%%%%%%%%
%


\section*{References}
\begin{thebibliography}{99}

\bibitem{ablt} I. Antoniadis, C. Bachas, D. Lewellen and T. Tomaras,
\Journal{\PLB}{207}{441}{1988}.

\bibitem{kp} C. Kounnas and M. Porrati, \Journal{\NPB}{310}{355}{1988};
S. Ferrara, C. Kounnas, M. Porrati and F. Zwirner,
\Journal{\NPB}{318}{75}{1989}.

\bibitem{ia} I. Antoniadis, \Journal{\PLB}{246}{377}{1990}.

\bibitem{tv} T. Taylor and G. Veneziano, \Journal{\PLB}{212}{147}{1988}.


\bibitem{ap} I. Antoniadis and B. Pioline, \Journal{\NPB}{550}{41}{1999},
hep-th/9902055.

\bibitem{w95} E. Witten, %\it Some comments on string dynamics},
Proceedings of Strings 95, hep-th/9507121; A.~Strominger,
\Journal{\PLB}{383}{44}{1996};{\bf B475}, 94 (1996).

\bibitem{sei} N. Seiberg, \Journal{\PLB}{390}{169}{1997}.

\bibitem{w} E. Witten, \Journal{\NPB}{471}{135}{1996}.

\bibitem{l} J.D. Lykken, \Journal{\PRD}{54}{3693}{1996}, 
G. Shiu and S.-H.H. Tye, \Journal{\PRD}{58}{106007}{1998}; 
Z. Kakushadze and S.-H.H. Tye, \Journal{\NPB}{548}{180}{1999}.

\bibitem{add} N. Arkani-Hamed, S. Dimopoulos and G. Dvali, 
\Journal{\PLB}{429}{263}{1998}.

\bibitem{aadd} I. Antoniadis, N. Arkani-Hamed, S. Dimopoulos and G. Dvali, 
\Journal{\PLB}{436}{263}{1998}. 
%
\bibitem{ab} 
I. Antoniadis, C. Bachas, \Journal{\PLB}{450}{99}{83}.
%
\bibitem{int}
K.~Benakli, \Journal{\PRD} {60}{104002}{1999}.

\bibitem{burg}
C.P. Burgess, L.E.
Ib{\'a}{\~n}ez,  F. Quevedo, \Journal{\PLB}{429}{257}{1999}.

\bibitem{price} See for instance: J.C. Long, H.W. Chan and J.C. Price,
\Journal{\NPB}{539}{23}{1999}.


\bibitem{hw} P. Ho\v{r}ava and E. Witten, 
%{\it Heterotic and Type I String Dynamics from Eleven Dimensions},
\Journal{\NPB}{460}{506}{1996}.

\bibitem{witten} E. Witten, \Journal {\NPB} {471}{135} {1996};




\bibitem{nse} K. Benakli,  \Journal{\PLB}{447}{51}{1999}; 
S. Stieberger, \Journal {\NPB} {541}{109} {1999};
Z. Lalak, S. Pokorski and S. Thomas, \Journal {\NPB} {549} {63}{1999}; 
D.G.~Cerde\~no and C.~Mu\~noz, \Journal{\PRD}{61}{016001}{2000}.


\bibitem{SO32} E. Witten, \Journal {\NPB} {460} {541}{1996}.



\bibitem{yaron} K.~Benakli and Y.~Oz, \Journal {\PLB}{472}{83}{2000}.


\bibitem{Kap} E. Caceres, V. S. Kaplunovsky and I. M. Mandelberg,
 {\em Nucl.Phys.}{\bf B493} 73.

\bibitem{aspin}  P. S. Aspinwall and D. R. Morrison, 
\Journal {\NPB} {503} {533}{1997}.


\bibitem{BI} J. D. Blum and  K. Intriligator, hep-th/9705044.


\bibitem{Intri} K. Intriligator, hep-th/9708117.

\bibitem{SW} N. Seiberg and  E. Witten,
{\em Nucl.Phys.} {\bf B471},  121 (1996).


\bibitem{GH}  O. Ganor and  A. Hanany,
{\em  Nucl.Phys.} {\bf  B474}, 122 (1996) .

\bibitem{aq} I. Antoniadis and M. Quir{\'o}s, 
%{\it Large Radii and String Unification},
\Journal{\PLB}{392}{61}{1997}.

\bibitem{ht} C.M. Hull and P.K. Townsend, 
%{\it Unity of Superstring Dualities},
\Journal{\NPB}{438}{109}{1995} and
\Journal{\NPB}{451}{525}{1995};
E. Witten, %{\it String theory dynamics in various dimensions},
\Journal{\NPB}{443}{85}{1995}.


\bibitem{ddg} K.R. Dienes, E. Dudas and T. Gherghetta, 
\Journal{\PLB}{436}{55}{1998};
\Journal{\NPB}{537}{47}{1999}.

\bibitem{poweruni} D. Ghilencea and G.G.
Ross, \Journal{\PLB}{442}{165}{1998}; Z. Kakushadze, 
\Journal{\NPB}{548}{205}{1999}; 
A. Delgado and M. Quir\'os, \Journal{\NPB}{559}{235}{1999}; 
P. Frampton and A. Rasin, \Journal{\PLB}{460}{313}{1999};
A. P\'erez-Lorenzana and R.N. Mohapatra, \Journal{\NPB}{559}{255}{1999};
Z. Kakushadze and T.R. Taylor, \Journal{\NPB}{562}{78}{1999}.


\bibitem{cb} C. Bachas, %{\it Unification with Low String Scale},
{\em JHEP} {\bf 9811}, 23 (1998).

\bibitem{admr} N. Arkani-Hamed, S. Dimopoulos and J. March-Russell,
hep-th/9908146.

\bibitem{abd} I. Antoniadis, C. Bachas and E. Dudas, 
\Journal{\NPB}{560}{93}{1999}.


\bibitem{ks} S. Kachru and E. Silverstein, {\em JHEP} {\bf 11}, 1 (1998),
hep-th/9810129; J. Harvey, \Journal{\PRD}{59}{26002}{1999}; R.
Blumenhagen and L. G{\"o}rlich, hep-th/9812158; C. Angelantonj, I.
Antoniadis and K. Foerger, \Journal{\NPB}{555}{116}{1999},
hep-th/9904092.

\bibitem{ads} I. Antoniadis, E. Dudas and A. Sagnotti, hep-th/9908023;
G. Aldazabal and A.M. Uranga, hep-th/9908072.

\bibitem{ba} C. Bachas, hep-th/9503030;
J.G. Russo and A.A. Tseytlin, \Journal{\NPB}{461}{131}{1996}.


\bibitem{ssopen} I. Antoniadis, E. Dudas and A. Sagnotti,
\Journal{\NPB}{544}{469}{1999}; I. Antoniadis, G. D'Appollonio, E. Dudas
and A. Sagnotti, \Journal{\NPB}{553}{133}{1999} and
hep-th/9907184.
%
\bibitem{gg} B. Grzadkowski and J.F. Gunion, \Journal{\PLB}{473}{50}{2000}.

\bibitem{abqhiggs} I.~Antoniadis, K.~Benakli and M.~Quir\'os, 
hep-ph/0004091.

\bibitem{SS} 
I.~Antoniadis, S.~Dimopoulos, A.~Pomarol and
M.~Quir{\'o}s, \Journal{\NPB}{544}{503}{1999};
%%CITATION = HEP-PH 9810410;%%
A.~Delgado, A.~Pomarol and
M.~Quir{\'o}s, \Journal{\PRD}{60}{095008}{1999}.
%%CITATION = HEP-PH 9812489;%%
%
\bibitem{higgs} J.A.~Casas, J.R.~Espinosa, M.~Quir{\'o}s and A.~Riotto,
\Journal{\NPB}{436}{3}{1995}; 
M.~Carena, J.R.~Espinosa, M.~Quir{\'o}s and
C.E.M.~Wagner, \Journal{\PLB}{355}{209}{1995};
M.~Carena, M.~Quir{\'o}s and
C.E.M.~Wagner, \Journal{\NPB}{461}{407}{1996};
H.E.~Haber, R.~Hempfling and A.H.~Hoang, \Journal{\ZPC}{75}{539}{1997};
M.~Carena, H.E.~Haber, S.~Heinemeyer, W.~Hollik, C.E.M.~Wagner and
G.~Weiglein, hep-ph/0001002;
J.R.~Espinosa and R.-J.~Zhang, {\em JHEP} {\bf 3}, 26 (2000).

%
\bibitem{adpq} I. Antoniadis, C. Mu\~noz and M. Quir\'os,
\Journal{\NPB}{397}{515}{1993}; A. Pomarol and M. Quir\'os,
\Journal{\PLB}{438}{225}{1998};
I. Antoniadis, S. Dimopoulos, A. Pomarol and M. Quir\'os,
\Journal{\NPB}{544}{503}{1999}; A. Delgado, A. Pomarol
and M. Quir\'os,  \Journal{\PRD}{60}{095008}{1999}.

\bibitem{AAB} E. Accomando, I. Antoniadis and  K. Benakli, 
\Journal{\NPB}{579}{3}{2000}.


\bibitem{ABQ} I.~Antoniadis and K.~Benakli, \Journal{\PLB}{326}{69}{1994},
I.~Antoniadis, K.~Benakli and M.~Quir\'os, \Journal{\PLB}{331}{313}{1994};
 P. Nath, Y. Yamada and M. Yamaguchi, \Journal{\PLB}{466}{100}{1999}
T.G. Rizzo and J.D. Wells,   \Journal{\PRD}{61}{016007}{2000};
A. De Rujula, A. Donini, M.B. Gavela and S. Rigolin, 
\Journal{\PLB}{482}{195}{2000}.

%

\bibitem{ABQ2}I.~Antoniadis, K.~Benakli and M.~Quir\'os, \Journal{\PLB}{460}
{176}{1999}.




\bibitem{Dlr} A.~Djouadi, A.~Leike, T.~Riemann, D.~Schaile and
C.~Verzegnassi, {\em Z. Phys.} {\bf C56} (1992) 289.
%





\bibitem{scc} P.~Nath and M.~Yamaguchi, 
\Journal{\PRD}{60}{116004}{1999}; \Journal{\PRD}{60}{116006}{1999};
M.~Masip and A.~Pomarol, \Journal{\PRD}{60}{096005}{1999}; 
W.J.~Marciano, \Journal{\PRD}{60}{093006}{1999};
A.~Strumia,  \Journal{\PLB}{466}{107}{1999};
R.~Casalbuoni, S.~De Curtis, D.~Dominici and R.~Gatto, 
\Journal{\PLB}{462}{48}{1999};
C.D.~Carone, \Journal{\PRD}{61}{015008}{2000}.
%
%
\bibitem{Delgado} A.~Delgado, A.~Pomarol and M.~Quir\'os, 
{\em JHEP} {\bf 1}, 30 (2000).

 
%

\bibitem{iadd} I. Antoniadis, S. Dimopoulos and G. Dvali,
\Journal{\NPB}{516}{70}{1998}.

%
\bibitem{missing} G.F. Giudice, R.~Rattazzi and J.D.~Wells,
\Journal {\NPB}{544}{3}{1999}; E.A.~Mirabelli, M.~Perelstein and M.E.~Peskin,
\Journal {\PRL}{82}{2236}{1999};  T.~Han, J.~D.~Lykken and R.~Zhang, 
 Phys.\ Rev.\  {\bf D59}, 105006 (1999); K. Cheung and W.-Y.
  Keung, Phys.\ Rev.\ {\bf D60}, 112003 (1999);  S.
  Cullen and M. Perelstein, Phys.\ Rev.\ Lett.\ {\bf 83} (1999) 268;
 C. Bal\'azs et al., Phys.\ Rev.\ Lett.\ {\bf 83}
  (1999) 2112;  L3 Collaboration (M. Acciarri et al.),
  Phys.\ Lett.\ {\bf B464}, 135 (1999), Phys.\ 
  Lett.\ {\bf B470}, 281 (1999); J.L.~Hewett, \Journal {\PRL}{82}{4765}{1999};
 D. Atwood, C.P. Burgess, E. Filotas, F. Leblond, D. London and I. Maksymyk, 
 hep-ph/0007178. For a recent analysis, see~\cite{string} and 
 references therein.


\bibitem{veneziano} G. Veneziano, \Journal{\NCA}{57}{190}{1968}; 
D. J. Gross and P. F. Mende, \Journal{\PLB}{197}{129}{1987}; 
\Journal{\NPB}{303}{407}{1988}; 
D. J. Gross and 
J.L. Manes, \Journal{\NPB}{326}{73}{1989}; S.H. Shenker, hep-th/9509132.

\bibitem{supernovae} S. Cullen and M. Perelstein, \Journal {\PRL}{83}{268}{1999}; V. Barger, T. Han, C. Kao and R.J. Zhang, \Journal {\PLB}{461}{34}{1999}.


\bibitem{COMPTEL} K.~Benakli and S.~Davidson, \Journal{\PRD}{60}{025004}{1999};
L.J. Hall and D. Smith, \Journal{\PRD}{60}{085008}{1999}. 

\bibitem{astcos} N. Arkani-Hamed, S. Dimopoulos and G. Dvali,
\Journal{\PRD}{59}{086004}{1999}.


\bibitem{fkz} S. Ferrara, C. Kounnas and F. Zwirner,
\Journal{\NPB}{429}{589}{1994}.






\bibitem{tvdil} T.R. Taylor and G. Veneziano,
\Journal{\PLB}{213}{450}{1988}.



\bibitem{string} S.~Cullen, M.~Perelstein and M.E.~Peskin, hep-ph/0001166; 
D. Bourilkov,  hep-ph/0002172; 
L3 Collaboration (M. Acciarri et al.), hep-ex/0005028.

\bibitem{RS} L. Randall,  R. Sundrum, \Journal{\PRL}{83}{4690}{1999}
\end{thebibliography}
\end{document}